\documentclass[preprint,nobibnotes,preprintnumbers,superscriptaddress,showpacs,sort&compress,longbibliography]{revtex4-1} 

\usepackage{graphicx}		
\usepackage{titlesec}
\usepackage{blindtext}
\usepackage{xfrac}		
\usepackage{subcaption}
\usepackage{multirow}	
\usepackage{gensymb}
\usepackage{amssymb}
\usepackage{balance}
\usepackage{soul}
\usepackage[normalem]{ulem}	
\usepackage{smartref}	
\usepackage{nomencl}
\usepackage[T1]{fontenc}
\makenomenclature

\usepackage[colorlinks=true,linkcolor=blue,citecolor=red,urlcolor=blue]{hyperref}		
																								
\newcommand{\comment}[1]{} %komentarz wielolinijkowy
\makeatletter
\renewcommand{\p@subsection}{}
\renewcommand{\p@subsubsection}{}
\makeatother

\usepackage{setspace}

\usepackage{etoolbox}
\renewcommand\nomgroup[1]{%
  \item[\bfseries
  \ifstrequal{#1}{A}{ }{%
  \ifstrequal{#1}{G}{Greek symbols}{%
  \ifstrequal{#1}{B}{Abbreviations }{}}}%
]}

\begin{document}																													 
%%%%%%%%%%%%%%%%%%%%%%%%%%%%																										 
%																																	 
%\preprint{}																 
																												
%%%%%%%%%%%%%%%%%%%%%%%%%%%%%%%%%%%%%%%%%%%%%%%%%%%%%%%%%%%%%%%%%%%%%%%%%%%%%%%%%%%%%%%%%%%%%%%%%%%%%%%%%%%%%%%%%%%%%%%%%%%%%%%%%%%%%%

\newcommand{\kvec}{\mbox{{\scriptsize {\bf k}}}}
\newcommand{\lvec}{\mbox{{\scriptsize {\bf l}}}}
\newcommand{\qvec}{\mbox{{\scriptsize {\bf q}}}}

\newcommand*{\PN}[1]{{\color[rgb]{0, 0, 1}{PN: #1}}}
\newcommand*{\AD}[1]{{\color[rgb]{0, 0.5, 0}{AD: #1}}}
\newcommand*{\WE}[1]{{\color[rgb]{1, 0, 0}{WE: #1}}}
\newcommand*{\AT}[1]{{\color[rgb]{0, 0.5, 0.5}{AT: #1}}}
\newcommand*{\PK}[1]{{\color[rgb]{0.949, 0.149, 0.835}{PK: #1}}}

%%%%%%%%%%%%%%%%%%%%%%%%%%%%
\def\eq#1{(\ref{#1})}
\def\sec#1{\hspace{1mm}section \ref{#1}}
\def\chap#1{\hspace{1mm}section \ref{#1}}
\def\fig#1{\hspace{1mm}Fig. \ref{#1}}
\def\figur#1{\hspace{1mm}Figure \ref{#1}}
\def\tab#1{\hspace{1mm}Table \ref{#1}}
%%%%%%%%%%%%%%%%%%%%%%%%%%%%
\title{Sensitivity analysis of wavy wall performance in turbulent separation control: Effects of amplitude and period variations}

%Estimation of the optimal amplitude and period of a wavy wall for controlling turbulent separation at $Re_{\tau}=2500$

%Estimation of the optimal amplitude and period of a wavy wall used as a control method of turbulent separation at $Re_{\tau}=2500$

%Estimating the optimal amplitude and period of a wavy wall for controlling turbulent separation at $Re_{\tau}=2500$

\author{Piotr Kamiński}
\affiliation{Department of Thermal Machinery, Czestochowa University of Technology, Al. Armii Krajowej 21, 42-200 Czestochowa, Poland}

\author{Witold Elsner}
\affiliation{Department of Thermal Machinery, Czestochowa University of Technology, Al. Armii Krajowej 21, 42-200 Czestochowa, Poland}

\author{Artur Tyliszczak}
\affiliation{Department of Thermal Machinery, Czestochowa University of Technology, Al. Armii Krajowej 21, 42-200 Czestochowa, Poland}

\author{Pawe{\l} Niegodajew} \email{pawel.niegodajew@pcz.pl}
\affiliation{Department of Thermal Machinery, Czestochowa University of Technology, Al. Armii Krajowej 21, 42-200 Czestochowa, Poland}

%%%%%%%%%%%%
%\date{\today} 
\date{August 20, 2024} 
\begin{abstract}
%%%%%%%%%%%%%%%%%%%%%%%%%%%%%%%%%%%%%%%%%%%%%
\footnotesize{The influence of transversely oriented sinusoidal wall corrugation on an incompressible isothermal flow in the near-wall region, subjected to adverse pressure gradient conditions at a friction Reynolds number of $Re_{\tau}=2500$, is investigated using Large Eddy Simulation. This study is a continuation of the work \cite{kaminski2024numerical} devoted to dynamics of the flow in the regions of crests and downhill/uphill waviness parts, and where the impact of the local and global pressure gradient were analysed. It focuses on the influence of the waviness parameters, such as amplitude ($A$) and number of waviness periods ($N_{\lambda}$) on the turbulent boundary layer separation. In terms of the effective slope, $ES\propto|{dA}/{dx}|$, where $x$ is the streamwise direction, the analyzed cases included the configurations with $ES$ ranging from $ES=0$ (flat wall) to $ES=0.2455$ ($A/\lambda=0.051$, where $\lambda$ is the length of a single waviness). 
Specifically, the primary objective was to identify a combination of $A$ and $N_{\lambda}$ that would result in the greatest increase in the wall-shear stress $\tau_w$, relative to the flat plate configuration, aiming to achieve the maximum postponement of turbulent separation.
It is shown that the waviness significantly affects the flow field by enhancing the turbulent kinetic energy that leads to increased $\tau_w$. In particular, it is shown that the enhancement in TKE depends mainly on $ES$ and not on particular values of $A$ and $N_{\lambda}$. Namely, the cases which are significantly diversified by $A$ and $N_{\lambda}$ but are characterised by the same $ES$ exhibit almost identical TKE profiles. The highest increase in the wall-shear stress compared to the reference flat plate configuration was observed for the combination of $A$ and $N_{\lambda}$ corresponding to $ES=0.1473$.}
\\

%\textbf{use your individual colours if you want to write something}

%\PN{zzzzzz}
%\WE{zzzzzz}
%\AT{zzzzzz}
%\PK{zzzzzz}

%

\end{abstract}
%\pacs{88.80.ff, 73.22.-f, 61.72.Ww}
\maketitle
%
%%%%%%%%%%%%%%%%%%%%%%%%%%%%%%%%%%%%%%%%%%%%%%%%%%%%%%%%%%%%%%%%%%%%%%%%%%%%%%%%%%%%%%%%%%%%%%%%%%%
%%\clearpage 
%%\textcolor{blue}{
%\begin{singlespace}
%%\begin{doublespace}
%%\mbox{}
%
%
%%\printnomenclature
%%\end{doublespace}
%\end{singlespace}
%%}
%%\clearpage

\clearpage
\section{Introduction} \label{intro}
%%%%%%%%%%%%%%%%%%%%%%%%%%%%%%%%%%%%%%%%%%%%%%%%%%%%%%%%%%%%%%%%%%%%%---------------------------------------------------------------------

The phenomenon of a turbulent flow being exposed to the adverse pressure gradient (APG) is a very common scenario that is seen in practice - the flow over the suction side of an airfoil \cite{lee2004investigation, tanarro2020effect, vila2020separating}, diffuser flows \cite{azad1996turbulent,apsley2000advanced,salehi2017computation,yadegari2020numerical} or flow around a turbine blade \cite{bons2005critical,bons2003effect,goyal2017experimental} are all well-known application examples. At strong APG conditions the turbulent boundary layer (TBL) separation \cite{nagano1993effects,kitsios2017direct,peterson2019control} may occur which leads to increased aerodynamic drag \cite{gad1991separation}. As such, a major decrease in the energy efficiency of a system takes place. The flow separation occurs when the wall-shear stress reaches zero ($\tau_w=\mu(dU/dy_w)=0$, where: $\mu$ is the dynamic viscosity and $dU/dy_w$ is the near-wall velocity gradient along the wall-normal direction $y_w$). Various techniques are available to control this phenomenon which are generally classified as active or passive \cite{joshi2016review,ashill2005review}. The first group involves systems, often quite complex, requiring an external energy source. Some examples include: synthetic jets \cite{you2008active}, blowing and suction \cite{moghaddam2017active}, heating and cooling \cite{jahanmiri2010active, yoon2006drag, harwigsson1996environmentally}, moving wall systems \cite{quadrio2011drag,leschziner2020friction}, electromagnetic lorentz force actuators \cite{jahanmiri2010active}, plasma actuators \cite{vernet2018flow, vernet2018plasma, vernet2015separation} and so on. Due to high costs, complexity and energy consumption of the active methods, the passive flow control is favoured as it does not require any external source of energy. The passive techniques involve some degree of surface modification or roughness and among these, the most common are for example: dimples on a golf ball \cite{lorenz2006spinning, choi2006mechanism, tay2018drag, tay2015mechanics, gattere2022dimples, aoki2012mechanism, bearman1976golf, veldhuis2009drag, tay2011determining, tay2019drag} or vortex generators on an airfoil \cite{koike2004research, aider2010drag, seshagiri2009effects}. Both of them are tools for flow turbulisation, postponement of separation and drag force reduction. Among unconventional flow control methods few of them are worth mentioning, i.e.: slotted airfoils \cite{belamadi2016aerodynamic, coder2020design, whitman2006experimental}, streamwise waviness for reducing the leading edge interaction noise \cite{casalino2019aeroacoustic, teruna2022numerical} or placing microcylinder near a blade leading edge \cite{WANG2018101, mostafa2022quantitative, wang2023wake}. There are also other solutions that are quite often nature-inspired. Some examples of these include: winglets \cite{smith2001performance,la2004induced,guerrero2012biomimetic,wu2018experimental,guerrero2020variable}, which are the wingtip devices often seen at the end of an airfoil and these are inspired by the ends of bird’s wing; fish scale microstructure, which are able to reduce drag at lower Reynolds numbers \cite{wu2018experimental, dou2012bionic};  herringbone riblets inspired by the bird’s feather \cite{chen2013biomimetic}; micro-scale riblets inspired by the skins of sharks \cite{schlieter2016mechanical} or dolphins \cite{lang2017separation} and so forth.  

The passive flow control methods, which were discussed so far, can employ either relatively simple or very complex surface modifications. These techniques are mostly applicable for laminar or transitional flows \cite{lissaman1983low, mueller2003aerodynamics} since their effectiveness diminishes with increasing $Re$ \cite{mcmasters1979low} and when applied to fully turbulent flows the separation occurs earlier. This might be attributed to the momentum deficit caused by the wall irregularities \cite{aubertine2004parameters,mejia2013wall}. 

In the literature, one can find several works devoted to the flow over the wavy wall surfaces conducted under ZPG conditions \cite{cherukat1998direct, sun2018direct, kuhn2010large, tyson2013numerical, kruse2006structure, hamed2015turbulent, elsner2022experimental, akselsen2020langmuir,fernex2020actuation, de1997direct, koyama2007turbulence, yoon2009effect, fujii2011turbulence, ghebali2017turbulent, segunda2018experimental, hamed2017turbulent}. A direct numerical simulation (DNS) of Angelis et al. \cite{de1997direct} has shown that for a certain amplitude range of a sinusoidal corrugation, the areas with increased wall-shear stress emerge. A flow field study over walls modified by waviness examined by Koyama et al. \cite{koyama2007turbulence} indicated enhancement of $\tau_w$ for one of the tested configurations. Yoon et al. \cite{yoon2009effect} showed that a corrugation with amplitude ($A$) to the period ($\lambda$) ratio $A/{\lambda}=0.03$ allows for a maximal enhancement of $\tau_w$ for Reynolds number based on the free-stream velocity $Re=(U_{\infty}H/\nu)=6760$ (where $U_{\infty}$ is the free-stream velocity, $H$ is the mean height of the channel and $\nu$ is the kinematic viscosity). The presence of wall waviness generally leads to the increase in velocity fluctuations and consequently to the increase in the turbulent kinetic energy (TKE) \cite{cherukat1998direct, sun2018direct, tyson2013numerical}. On the other hand, when increasing $A/\lambda$, an enhancement of the aerodynamic drag as well as the decrease in wall-shear stress is observed \cite{fujii2011turbulence, ghebali2017turbulent, segunda2018experimental, hamed2017turbulent}. 

Enhancement of $\tau_w$ due to the wavy wall in some of the mentioned studies, although obtained for ZPG flows, triggered the research group of Dr{\' o}{\. z}d{\. z} et al. \cite{drozdz2021effective} to verify whether the effect can be obtained under APG conditions. In particular, the authors examined experimentally the usefulness of the surface undulation in the enhancement of the wall-shear and postponement of the turbulent separation at high $Re$.
It was shown that for a certain amplitude and period of a transverse wall waviness, a 13\% increase in skin-friction can be obtained at Reynolds number defined based on the boundary layer thickness $\delta$ and the friction velocity $u_\tau=\sqrt{\tau_w/\rho}$ (where $\rho$ is the mass density) equal to $Re_{\tau}=u_\tau \delta/\nu=4000$. This allows to move significantly the point of turbulent separation downstream. They also showed that the mechanisms governing the increase in $\tau_w$ caused by the wavy wall are similar to the amplitude modulation of small scales by the large scales \cite{mathis2009comparison, dogan2019quantification, andreolli2023separating}. 

In our previous work \cite{elsner2022experimental} we examined, experimentally and numerically, the effects of different surface topology (two-dimensional and three-dimensional) on the turbulent flow development over a smooth wall downstream the corrugation at $Re_{\tau}=1350$ under APG conditions. It was found that for this particular $Re_{\tau}$ value, none of the investigated surface topologies ensured an enhancement in the wall-shear stress in the region of interest (i.e. downstream the corrugation) with respect to the flat plate case. The lack of enhancement in $\tau_w$ should be attributed to the insufficiently high value of $Re_{\tau}$ examined. It is because, in the most recent work of Kamiński et al. \cite{kaminski2024numerical} exploring the flow over a wavy wall under zero and adverse pressure gradient conditions indicated that enhancement in $\tau_w$ and so, a postponement of TBL separation, is possible at higher $Re_{\tau}$ of 2500. 

This work further extends the research of Kamiński et al.~\cite{kaminski2024numerical} by a systematic search for the values of corrugated wall parameters, i.e., the amplitude ($A$) and a number of periods ($N_{\lambda}$), allowing for a maximum increase in $\tau_w$ and also further examination of the impact of these parameters on the flow picture under APG conditions. The analysis is conducted with the use of the Large Eddy Simulations (LES) in ANSYS Fluent software for the same Reynolds number as in \cite{kaminski2024numerical} equal to $Re_{\tau}=2500$ corresponding to Reynolds number defined based on the momentum thickness $\theta=\int_0^{\delta} \langle U(y)\rangle/U_{\infty})(1-\langle U(y)\rangle/U_{\infty})dy$ equal to $Re_{\theta}=U_{\infty}\theta/\nu=9800$. The model's performance was verified in our previous work \cite{kaminski2024numerical} where an identical numerical setup was employed. Now, the focus is put on the near-wall turbulence within the waviness as well as downstream within the flat plate region. In particular, it is investigated with LES whether the transverse corrugation has a long-lasting effect on the near-wall flow downstream. The key goal is to find an optimal combination of $A$ and $N_{\lambda}$, allowing for the maximum increase in skin-friction that should ensure the maximum postponement of TBL detachment. An important outcome of the research is a better understanding of the physics of the flow modified by the corrugation that may be beneficial for further wall topology optimisation under different flow conditions.

The present manuscript is organised as follows: the description of the research object and numerical methods, together with details of the wavy wall configurations, as well as mesh characteristics, are described in Section \ref{numerical-modeling}; the results obtained using LES are presented in Section \ref{results}; and the conclusions and suggestions for future work are provided in Section \ref{conclusions}.

\newpage
\section{Methods \label{numerical-modeling}}
\subsection{Research object}

This research examines an incompressible and isothermal TBL developing over a 2D streamwise wavy wall subjected to APG conditions at $Re_{\tau}=2500$ using LES. In particular, the present work extends the previous one \cite{kaminski2024numerical} by providing a detailed systematic study aimed at distinguishing the most effective combination of amplitude and period of the waviness in postponement of turbulent separation. 

The waviness is defined using the sine function parametrised by $A$ and ${\lambda}$ according to the relation: 
\begin{equation}
       y_w(x)=A\cdot \sin(\omega x  )
\end{equation}
The reference shape reflects the one analysed in~\cite{kaminski2024numerical} and is characterised by the amplitude defined as:
\begin{equation}
    A= A_0 (0.00366x^2 + 0.000614x + 0.003351)
\end{equation}
The parameter $A_0$ is the dimensionless amplitude scale that allows for increasing or decreasing the amplitude with respect to the basic one $A_0=1.0$ that was examined in \cite{kaminski2024numerical} for identical $Re_{\tau}=2500$. The variation of the amplitude in the streamwise direction ensured that $A$ increases proportionally to $\delta$ and barely touched the outer region of TBL as suggested in \cite{drozdz2021effective}. The frequency of the waviness is defined as $\omega=2\pi/\lambda$ with $\lambda=L_{x,w}/N_{\lambda}$, which for the reference case with $N_{\lambda}=5$ equals $\lambda=0.1332$~m. Cases with (i) the constant period number $N_{\lambda}=5$ and varying amplitude in the range $A_0=0-2.0$; (ii) the cases with the constant amplitude $A_0=1.0$ and varying periods number $N_{\lambda}=1-7$ were considered. Note, that the case with $A_0=0$ corresponds to the configuration with the flat wall. 

\begin{figure}[h!] 
    \centering
       \includegraphics[trim={0 0 0 0}, clip, width=0.9\linewidth]{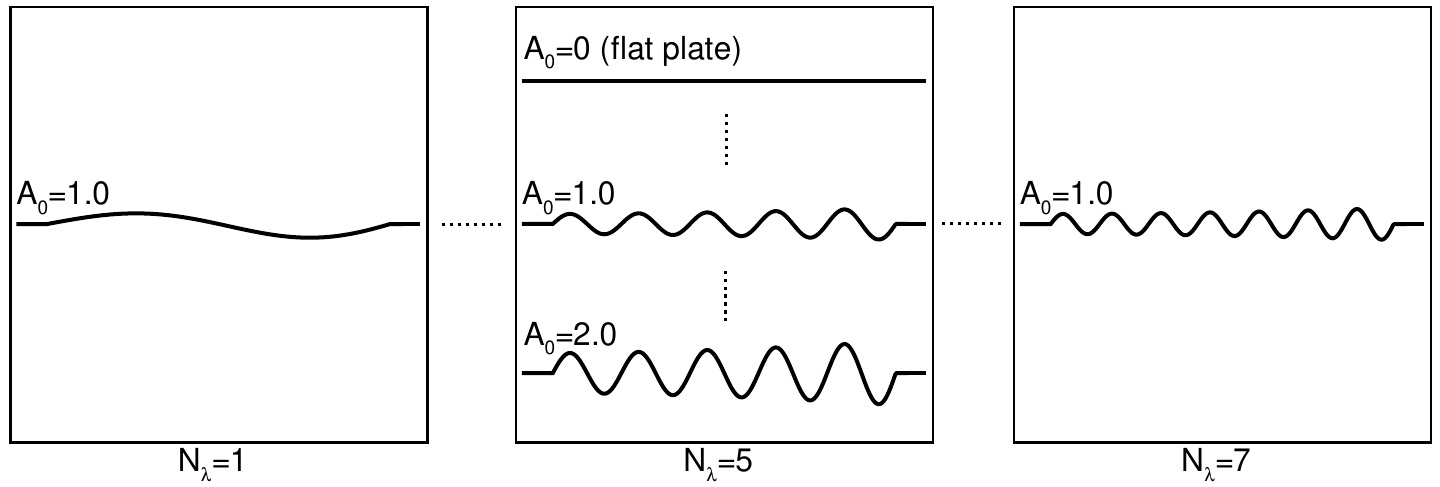}
    \caption{Schematic drawing illustrating corrugation geometries described in Tables \ref{Tab:Adiff} and \ref{Tab:NLdiff}.}
    \label{fig:geometries-schematic}
\end{figure}

Figure \ref{fig:geometries-schematic} presents the schematic drawing of selected surfaces considered in the study, involving changing $A_0$ and $N_{\lambda}$, whereas Tables~\ref{Tab:Adiff} and \ref{Tab:NLdiff} report the main parameters of all test cases investigated, such as the amplitude $A$ at the localisation of the tip of the first period ($x=0.043$ m), the amplitude expressed in viscous units, i.e. $A^+ = A\cdot u_{\tau}(x)/\nu$ (where $u_{\tau}(x)$ is measured on the flat wall) and averaged across the wavy wall, the ratio $A(x=0.043$ m$)/{\lambda}$ and the effective slope defined as the mean absolute streamwise gradient of the corrugation ($ES = \frac{1}{L_{x,w}}\int_{L_{x,w}}|\frac{dA}{dx}|dx$) \cite{napoli2008effect} for particular combinations of $A_0$ and $N_{\lambda}$. The wavelength to period ratio for $N_{\lambda}=5$ (Tab. \ref{Tab:NLdiff}) is $\lambda/\delta_{avg}\approx 1.25$. The boundary layer thickness used for calculating $\lambda/\delta_{avg}$ is taken as an average $\delta$ measured at $0.01$ m$<x<0.676$ m for the flat wall case ($A_0=0$), i.e. at the exact location of the corrugation (see Fig.\ref{fig:channel-1a} illustrating a schematic view of the experimental setup from Ref. \cite{kaminski2024numerical}). Note, that since $A$ grows along the $x$ direction, averaged value of $ES$ along whole wavy wall is given instead of the one for a single period. According to the literature \cite{napoli2008effect, schultz2009turbulent,nugroho_monty_utama_ganapathisubramani_hutchins_2021}, based on the value of $ES$, the surface roughness can be classified as: waviness region for $ES<0.15$; transition region for $0.15<ES<0.35$; roughness regime for $ES>0.35$. In the waviness regime the viscous drag is dominant while in the rough regime the dependence of $U^+=U/u_{\tau}$ on $ES$ weakens. As shown by De Marchis \cite{de2016large} and Forooghi et al. \cite{forooghi2017toward}, for corrugated walls, $ES$ is regarded as the key factor controlling the near-wall flow, while for rough surfaces, the flow depends on the height of the roughness rather than on $ES$. Since the values of $ES$ in the present study are within the range $0\leq ES \leq 0.2455$ for all cases considered, the investigated surfaces can fit into the waviness or transition regime. 

\newpage
\begin{table}[ht]
%\begin{tabular}{|c|c|c|c|c|c|c|}
\begin{tabular}{p{30pt}p{50pt}p{40pt}p{40pt}p{30pt}p{40pt}}
\hline\hline
{$A_0$} & $A$~[m]  & ${A/{\lambda}}$ & $ES$ & ${A^+}$ & $A/{\delta}$\\ \hline
0 & 0& 0& 0& 0 &   0    \\ %\cline{1-2} \cline{4-6} 
0.2& 0.00068& 0.0051       & 0.0245& 23  & 0.0083\\% \cline{1-2} \cline{4-6} 
0.4& 0.00136& 0.0102       & 0.0491& 46     & 0.0166     \\% \cline{1-2} \cline{4-6} 
0.6& 0.00204& 0.0153       & 0.0736& 69   & 0.0249       \\% \cline{1-2} \cline{4-6} 
0.8& 0.00272& 0.0204       & 0.0982& 92    & 0.0332      \\% \cline{1-2} \cline{4-6} 
0.9& 0.00306& 0.0230        & 0.1105& 104  &    0.0373    \\ %\cline{1-2} \cline{4-6} 
1.0& 0.00340& 0.0255       & 0.1227& 115   & 0.0414      \\ %\cline{1-2} \cline{4-6} 
1.1& 0.00374& 0.0280       & 0.1350& 127  & 0.0456       \\% \cline{1-2} \cline{4-6} 
1.2& 0.00408& 0.0306       & 0.1473& 139  & 0.0497       \\% \cline{1-2} \cline{4-6} 
1.3& 0.00442& 0.0331       & 0.1595& 150  & 0.0539       \\ %\cline{1-2} \cline{4-6} 
1.4& 0.00476& 0.0357       & 0.1719& 162  & 0.0580       \\ %\cline{1-2} \cline{4-6} 
1.5& 0.00510& 0.0382       & 0.1841& 157  & 0.0622       \\% \cline{1-2} \cline{4-6} 
1.6& 0.00544& 0.0408       & 0.1964& 183   & 0.0663      \\% \cline{1-2} \cline{4-6} 
1.8& 0.00612& 0.0459       & 0.2209& 207   & 0.0746      \\ %\cline{1-2} \cline{4-6} 
2.0& 0.00680& 0.0510        & 0.2455& 231  & 0.0829       \\ \hline \hline
\end{tabular}
\caption{Test cases with the constant $\lambda=0.133\ m$ ($\lambda/\delta_{avg}\approx 1.25$) and varying amplitude. The amplitude $A$ [m] and $\delta$ (used for calculating $A/\delta$) are measured at the tip of the first wave, i.e. at $x=0.043$ m.}
\label{Tab:Adiff}
\end{table}

\begin{table}[ht]
%\begin{tabular}{|c|c|c|c|}
\begin{tabular}{p{30pt}p{50pt}p{50pt}p{40pt}p{40pt}}
\hline
\hline
$N_{\lambda}$ & ${\lambda}$ {[}m{]} & $\lambda/\delta_{avg}$ & ${A/{\lambda}}$ & $ES$  \\ \hline
1  & 0.666  & 6.25 & 0.0051  & 0.0244  \\ 
2  & 0.333  &3.12 & 0.0102  & 0.0491 \\ 
3  & 0.222  &2.08 & 0.0153  & 0.0736 \\ 
4  & 0.166  &1.56  & 0.0204  & 0.0982  \\ 
5  & 0.133  &1.25   & 0.0255    & 0.1227   \\ 
6  & 0.111  &1.04 & 0.0306   & 0.1473   \\ 
7  & 0.095 &0.89&   0.0357  &   0.1719  \\
\hline\hline
\end{tabular}
\caption{Test cases with the constant $A_0=1.0$ and varying number of periods. The TBL thickness $\delta$ in 3$^{rd}$ column was taken as the average value over the streamwise corrugation length $\delta_{avg}$.}
\label{Tab:NLdiff}
\end{table}

\subsection{Numerical model}

The LES method has been developed for over 60 years, leading to its current maturity and perception as a reliable CFD tool~\cite{Georgiadis2010,Zhiyin2015,Geurts2019}.
In this research, the computations were performed with the help of well-known and widely used ANSYS Fluent software applying the 2$^{nd}$ order bounded central differencing scheme for the spatial discretisation of the convective terms of the Navier-Stokes equations and 2$^{nd}$ order central scheme for the viscous terms. The SIMPLE algorithm with the 2$^{nd}$ order discretisation of the pressure gradient was used for the pressure-velocity coupling. The time integration method was performed by applying the 2$^{nd}$ order implicit method. The sub-grid scale tensor was modelled using the WALE model~\cite{nicoud1999subgrid}, which is dedicated especially for wall-bounded flows. In all analyzed cases the initial velocity in the computational domain was set to zero. A transient phase, until a fully turbulent flow was developed, lasted for approximately $0.27$~s, which corresponds to $3$ flow-through time units defined as $t_{FT}=L_x/U_{\infty}=0.09$~s. The time-averaging procedure lasted for the next 40 $t_{FT}$. Additionally, the solutions were spatially averaged in the $z$-direction by collecting and averaging data extracted from $10$ uniformly spaced cross-sections. This led to a good convergence of the time-averaged solutions, which was checked by comparing the results at the time instants  $38t_{FT}$ and $40t_{FT}$.  
%Depending on the mesh density and time-step, a single run required approximately 20 days of continuous computations on a computer cluster with $96$ CPUs.

\subsection{Numerical domain and boundary conditions}
\begin{figure}[ht]
\begin{minipage}[c]{0.80\linewidth}
\includegraphics[width=\linewidth]{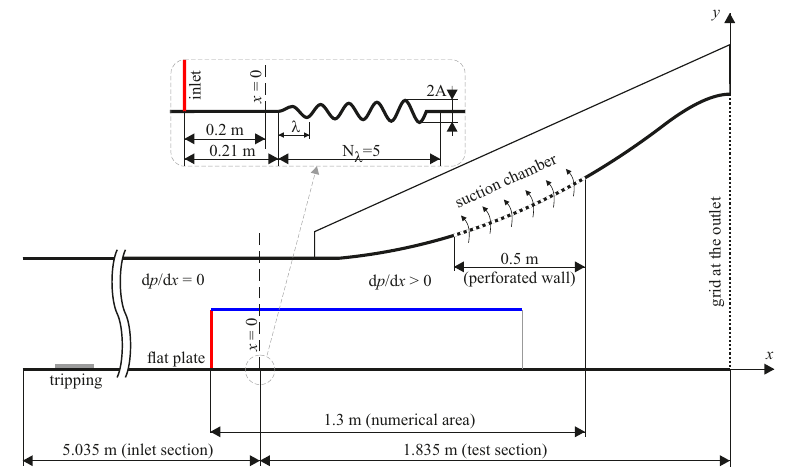}
\caption{The geometry from the experiment.}
\label{fig:channel-1a}
\end{minipage}
\end{figure}

\begin{figure}[ht]
\begin{minipage}[c]{0.8\linewidth}
\includegraphics[width=\linewidth]{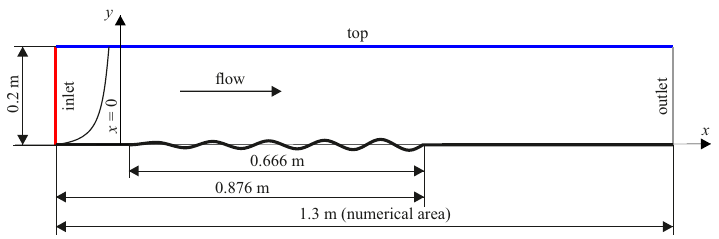}
\caption{Computational domain for LES study.}
\label{fig:channel-1b}
\end{minipage}
\end{figure}

LES allows for deep and detailed insights into flow dynamics, but its computational expense can make its application in parametric or optimization-oriented studies prohibitive. Therefore, for the present investigations only the central part of the experimental stand \cite{kaminski2024numerical,drozdz2021effective}, indicated by the red, grey and blue lines in Fig.~\ref{fig:channel-1a}, was considered in the simulations.  Figure~\ref{fig:channel-1b} shows the localisation of the wavy wall and the boundary conditions in the computational domain. Its dimensions were $L_x=1.3$~m ($L_x/\delta_{in}=16.17$, where $\delta_{in}$ is the inlet boundary layer thickness) in the streamwise direction, $L_y=0.2$~m in the wall-normal direction, and $L_z=0.24$~m in the spanwise $z$-direction. Despite this simplification, the simulations remained relatively time-consuming. A single run, which included both the initial transient phase and the time required to obtain convergent statistics, lasted for 20 days of continuous computations on a computer cluster equipped with 96 CPUs.

The assumed rectangular computational geometry required a prescription of the boundary conditions on its inlet and the top side reflecting precisely the experimental data. To do that, the measured velocity values in these localisations were approximated by high-order polynomials. On the inlet side, the streamwise velocity $\langle U_x\rangle$ and TKE profiles were defined. The former corresponded to the measured time-averaged data while the TKE profile served as a mask imposed on instantaneous turbulent velocity components generated by applying the synthetic turbulence generator method \cite{fluent-theory-guide}. This profile was defined such that $\langle u_xu_x\rangle^{1/2}$ obtained in the simulations agreed almost exactly with the measurements. 

\begin{figure}[ht] 
    \centering
    \begin{subfigure}{0.45\linewidth}
    \includegraphics[width=\linewidth]{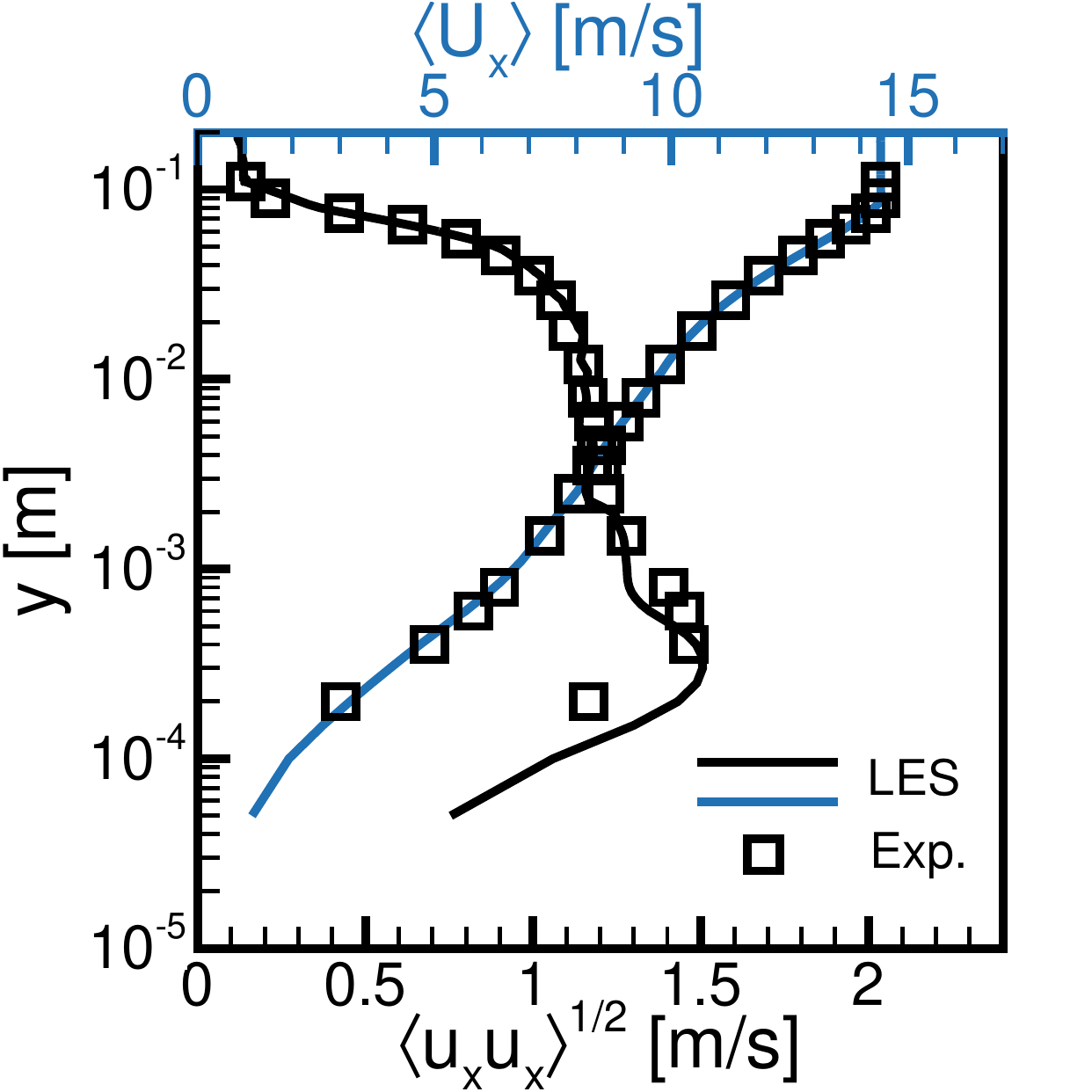}
    \caption{}
    \end{subfigure}
    \begin{subfigure}{0.45\linewidth}
    \includegraphics[width=\linewidth]{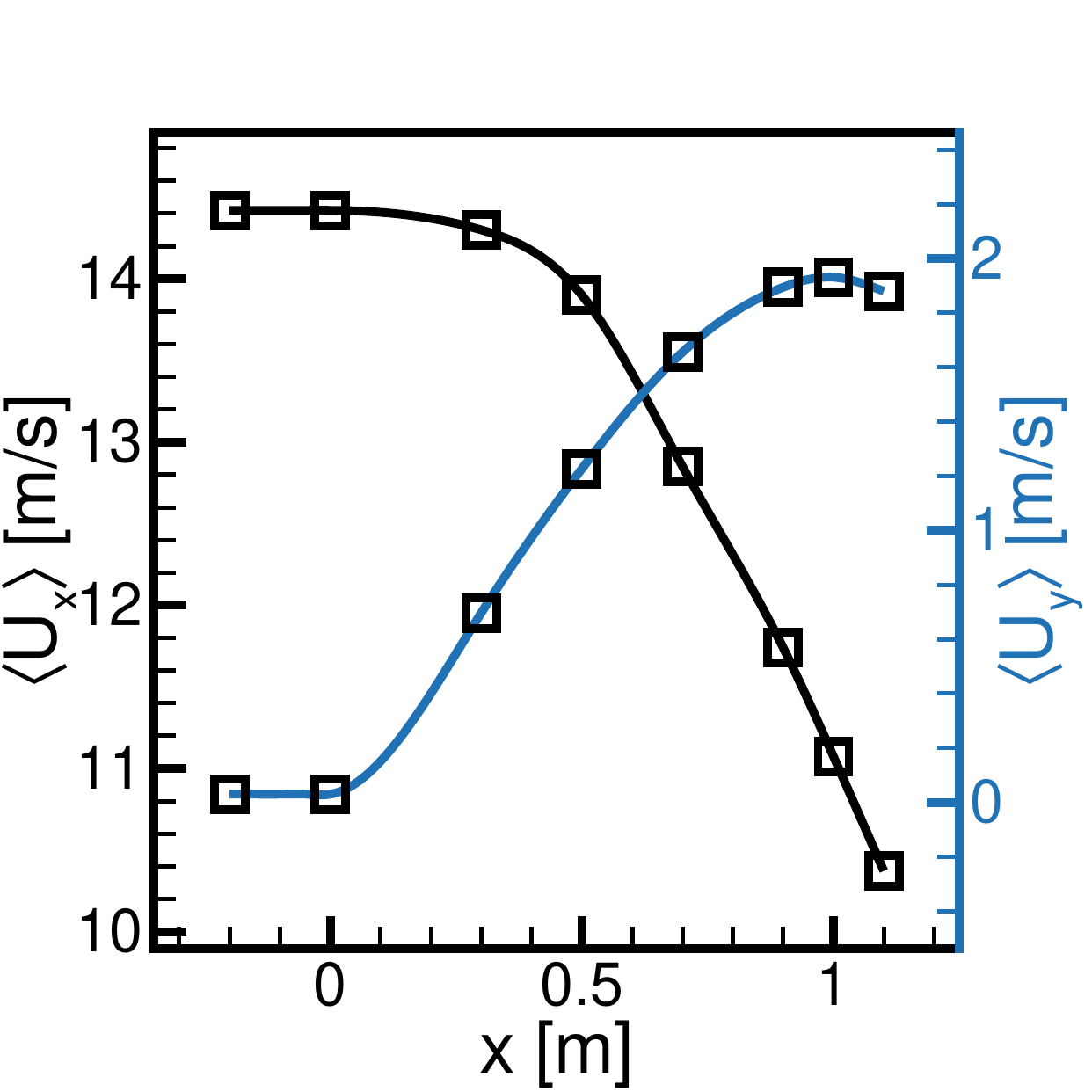}
    \caption{}
    \end{subfigure}
    \caption{The approximated profiles: for the inlet boundary (a) and for the top boundary (b) generating the APG condition.}
    \label{fig:inlet-conditions}
\end{figure}

Figure~\ref{fig:inlet-conditions} shows the profiles of the mean streamwise velocity and its fluctuation which were imposed on the inlet (Fig. \ref{fig:inlet-conditions}a). On the top boundary (Fig. \ref{fig:inlet-conditions}b), the velocity fluctuations were negligibly small, as will be shown later, and therefore only the mean streamwise $\langle U_x\rangle$ and wall-normal $\langle U_y\rangle$ velocity components were assumed in this localisation. On the side boundaries, spaced by $z=0.24$ m, the periodic boundary conditions were applied. This allowed  eliminating the need to refine the computational mesh near these boundaries that would be necessary if they were treated as non-slip walls. The test computations showed that the assumed periodicity has no impact on the solution obtained in the central part of the domain. The bottom wall of the domain was treated as the non-slip wall. At the outlet, the constant atmospheric pressure ($101325$ Pa) was assumed. The computational configuration described above closely corresponds to the one used in \cite{kaminski2024numerical}. In that study, the correctness of the applied simplifications of the computational geometry and prescribed boundary conditions was thoroughly verified. A high accuracy of the LES method was confirmed by comparing the computed and measured velocity and fluctuation profiles at selected locations above the waviness. We will not repeat these comparisons here and refer the interested reader to Ref. \cite{kaminski2024numerical}. 

\begin{figure}[h!] 
    \centering
       \includegraphics[trim={0 0 0 0}, clip, width=0.65\linewidth]{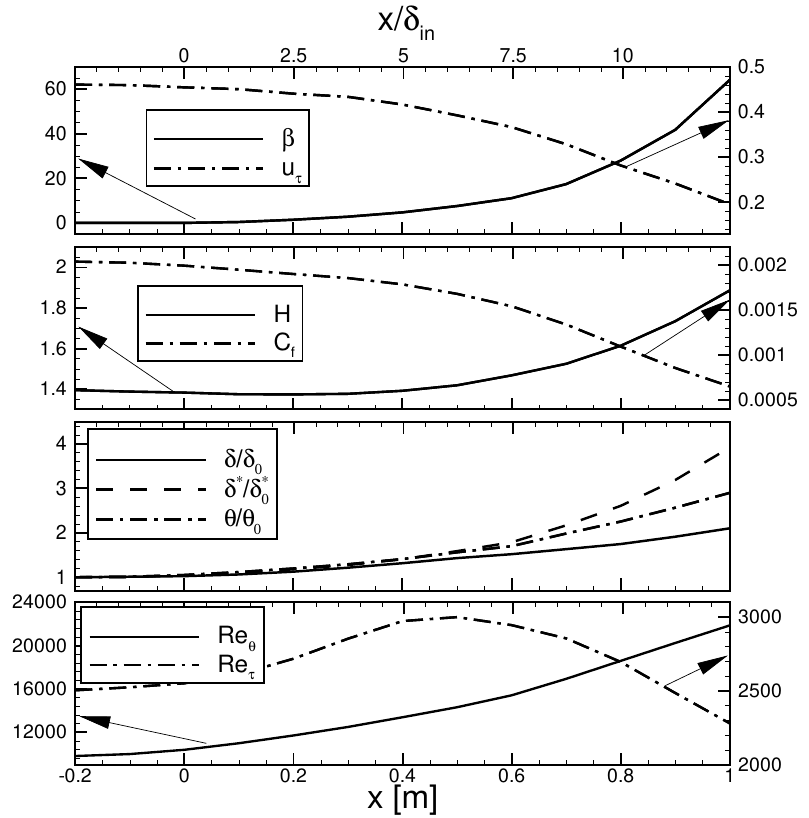}
    \caption{Main parameters of the turbulent boundary layer. The values of $\delta,\,\delta^*,\, \theta$ are normalized by their initial values at $x_0$.}
    \label{fig:flow-characteristics}
\end{figure}

Figure~\ref{fig:flow-characteristics} presents the profiles of characteristic TBL quantities along the domain for the case with the flat bottom wall, including the Clauser-Rotta pressure-gradient parameter $\beta = {\delta^*}/{\tau_w}\cdot{dP_{\infty}}/{dx}$ \cite{clauser1956turbulent} (where $\delta^*=\int_0^{\delta} (1-\langle U(y)\rangle/U_{\infty})dy$ is the displacement thickness), $u_\tau$ and the shape factor $H=\delta/\theta$%(where $\theta=\int_0^{\delta} \langle U(y)\rangle/U_{\infty})(1-\langle U(y)\rangle/U_{\infty})dy$ is the momentum thickness)
, skin-friction coefficient ($C_f=2(u_\tau/U_{\infty})^2$), and the Reynolds numbers calculated based on $\theta$ and $\delta$. The $x-$axis, representing the streamwise coordinate, was also provided in the non-dimensional form, i.e., normalised by $\delta_{in}$. The $\beta$ distribution shows that up to $x=0.3\ m$, the flow can be regarded as ZPG and the APG section starts downstream. This causes $\delta$, $\delta^*$ and $\theta$ to grow proportionally with $\beta$. Thus, enhancement of $H$ is also observed. As a result, $u_{\tau}$, $C_f$ and $Re_{\tau}$ decrease respectively, which indicates that the flow approaches separation.

\subsection{Computational mesh}

\begin{figure}[ht!] 
    \centering
     \begin{subfigure}{0.495\linewidth}
    \includegraphics[width=\textwidth]{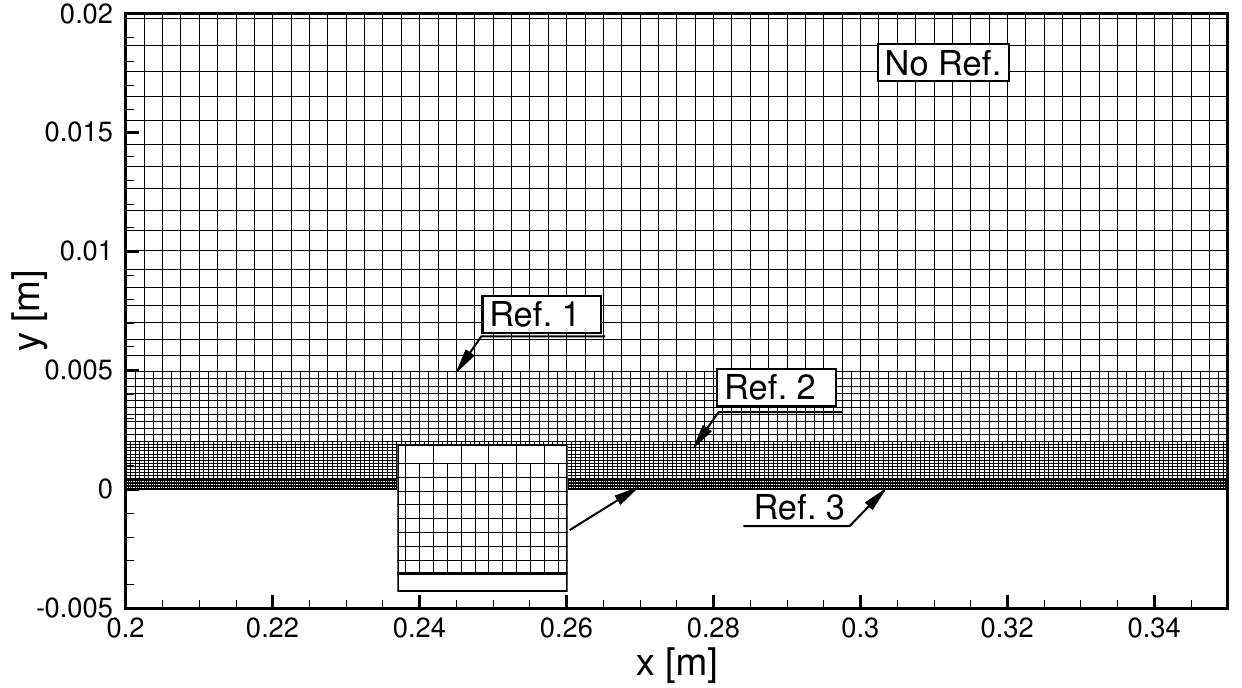}
    \caption{}
    \end{subfigure}
    %%%
     \begin{subfigure}{0.495\linewidth}
    \includegraphics[width=\textwidth]{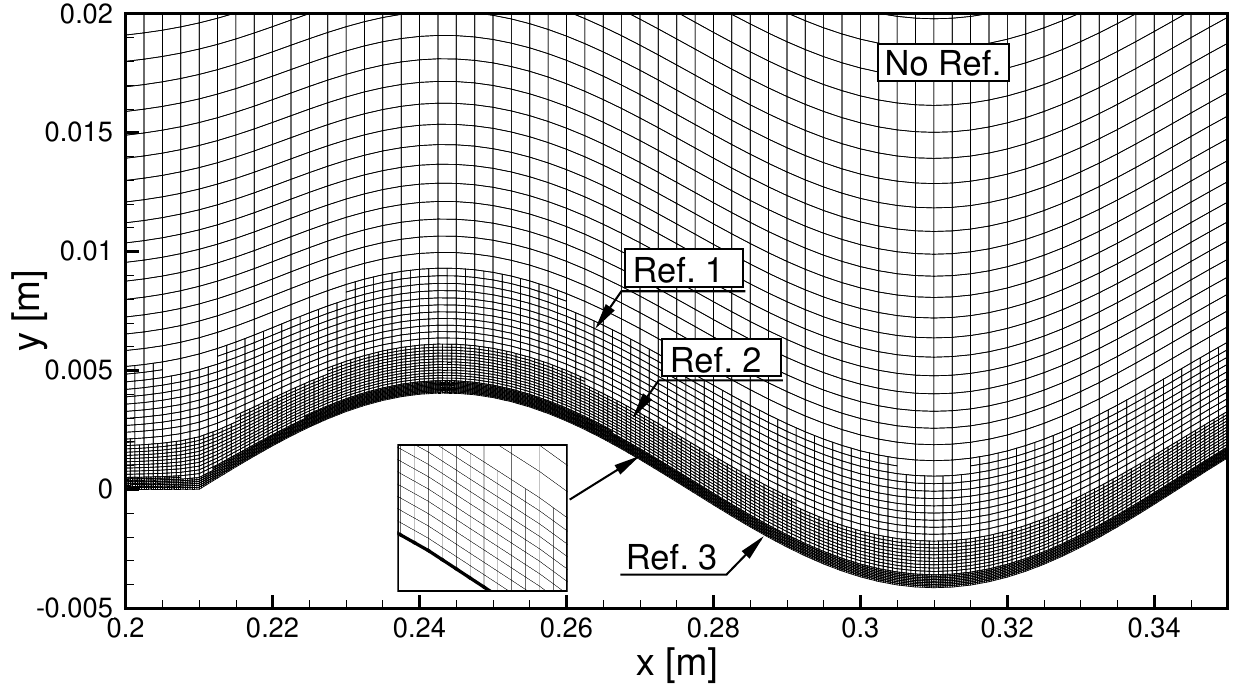}
    \caption{}
    \end{subfigure}
      \caption{Computational mesh M4 for cases $A_0=0$ (a) and $A_0=1.0$, $N_{\lambda}=5$ (b) used in the simulations.}
    \label{fig:meshes}
\end{figure}

The impact of mesh density on solution accuracy was analysed in \cite{kaminski2024numerical}, where solutions obtained on four block-structured meshes (M1-M4, see Table~\ref{Tab:Mesh_details}) were compared. These meshes differed in near-wall refinement, as illustrated in Fig.\ref{fig:meshes}. Specifically, Fig.\ref{fig:meshes} depicts mesh M4 for configurations with both the flat wall ($A_0=0$) and the wavy wall with $A_0=1.0$ and $N_{\lambda}=5$. Mesh M4 is composed of a one-block structural mesh M1, which was then refined near the wall. The first level of refinement, denoted as Ref. 1, resulted in mesh M2. Subsequent refinements, Ref. 2 and Ref. 3, led to meshes M3 and M4. Each refinement involved dividing a single cell into eight smaller cells. This strategy is useful for increasing mesh resolution near the wall while maintaining coarser mesh in regions with smaller velocity gradients in the upper part of the boundary layer and the free-stream flow. A similar mesh generation procedure was applied in the LES of the NACA4412 airfoil \cite{vinuesa2021high}.

Table~\ref{Tab:Mesh_details} presents the main parameters of all four meshes, including the minimum and maximum cell sizes neighbouring the wall expressed in viscous units ($y^+$, $x^+$, $z^+$). These values were calculated after performing computations on the respective meshes. It is worth noting that the spanwise and streamwise cell sizes ($\Delta z$, $\Delta x$) were the same, ensuring that the aspect ratios of the near-wall cells ($AR_w = \Delta x/\Delta y = \Delta z/\Delta y$) remained the same across all meshes. In \cite{kaminski2024numerical}, it was shown that the solutions obtained on meshes M3 and M4 were almost identical. However, for the present studies, similarly as in \cite{kaminski2024numerical}, the mesh M4 was selected. On this mesh, the condition $y^+<1$, which is regarded as necessary for LES of wall-bounded flows \cite{pope2001turbulent}, was fulfilled for all near-wall cells. Additionally, according to Pope \cite{pope2001turbulent}, to consider LES well resolved, the computational mesh should ensure that the ratio of sub-grid to total TKE, expressed as $k_{sgs}/k_{tot}$ must be below 20\%. In the case of mesh M4 this ratio was below 1\%. The computations were performed with the time-step $\Delta t=1.0\times 10^{-4}$~s that on the mesh M4 led to $CFL\approx0.8$. Comparisons of the solutions obtained with the $\Delta t=2.0\times 10^{-4}$~s did not reveal any significant differences, see also \cite{kaminski2024numerical}. 

\begin{table}[ht]
\begin{tabular}{p{40pt}p{70pt}p{70pt}p{70pt}p{70pt}p{40pt}}
\hline
\hline
Mesh & No. of cells & $y^+_\textrm{\tiny MIN,MAX}$ & $x^+_\textrm{\tiny MIN,MAX}$ & $z^+_\textrm{\tiny MIN,MAX}$ & $AR_w$ \\
\hline
M1 & $4.0\times 10^6$ & 1.98/6.61 & 20.68/68.85 & 20.68/68.85 & 5.2 \\
M2 & $7.2\times 10^6$ & 1.05/3.54 & 10.89/36.88 & 10.89/36.88 & 5.2 \\
M3 & $18.3\times 10^6$ & 0.52/1.88 & 5.47/19.56 & 5.47/19.56 & 5.2  \\
M4 & $41.1\times 10^6$ & 0.29/0.92 & 3.07/9.43 & 3.07/9.43 & 5.2  \\
\hline\hline
\end{tabular}
\caption{Parameters of the computational meshes. The cell sizes at the wall are
given in viscous units.}
\label{Tab:Mesh_details}
\end{table}

\section{Results \label{results}}

In the following sections, we discuss the results obtained on the densest mesh M4 with $\Delta t=1.0\times 10^{-4}$~s. As the complexity of the analysed cases does not differ from those for which the accuracy assessment was carried out the previous work \cite{kaminski2024numerical}, one can assume that the obtained results are accurate and reliable to a comparable extent. 

\begin{figure}[ht]
    \centering
    \includegraphics[width=0.8\textwidth]{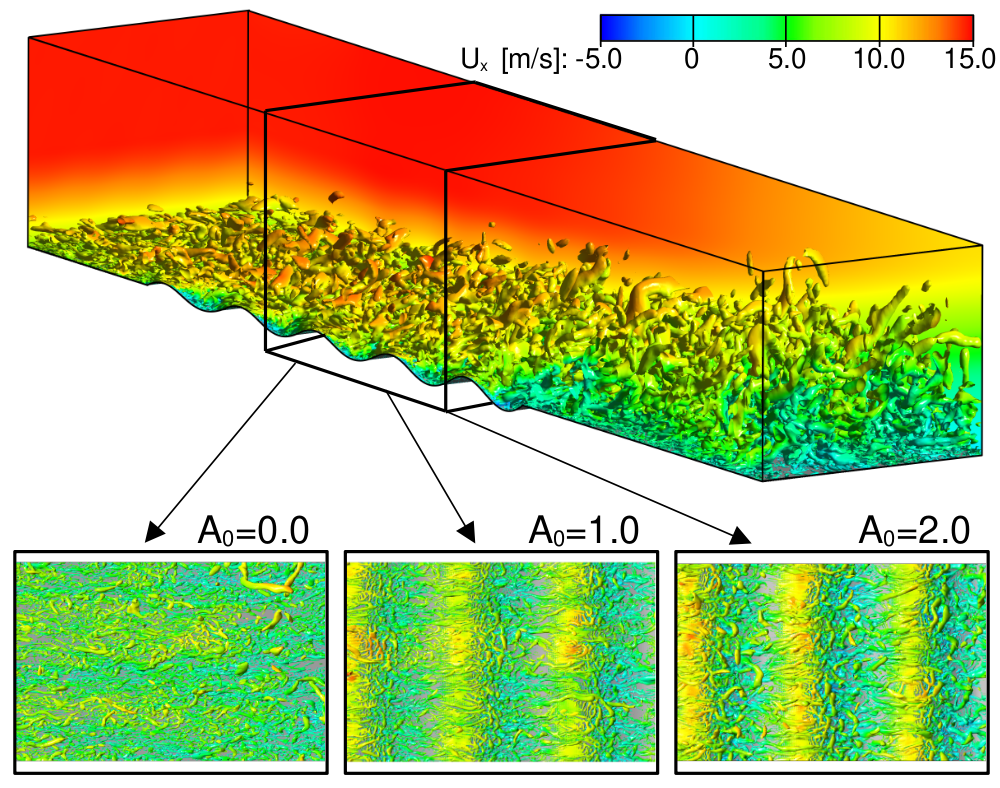}
    \caption{Q--parameter isosurface ($Q=0.05$~s$^{-2}$) coloured by the streamwise velocity shown on the back side of the computational geometry.}
    \label{fig:q-param}
\end{figure}

Figure~\ref{fig:q-param} shows a sample view of instantaneous values of the $Q$--parameter coloured by the streamwise velocity component for the cases with $A_0=0$, $A_0=1.0$ and $A_0=2.0$. The $Q$--parameter is defined as  $Q = 0.5(\Omega_{ij} \Omega_{ij} - S_{ij} S_{ij})$, where $\Omega_{ij}$ and $S_{ij}$ are antisymmetric and symmetric velocity gradient tensors. Positive values of the $Q$--parameter are often used as indicators of coherent vortical structures~\cite{hunt1988eddies}. Here, the $Q$--parameter ($Q=0.05$~s$^{-2}$) nicely visualizes the complexity of the flow and the presence of near-wall structures. In the case with $A_0=0$, structures are evidently smaller and more irregular than for $A_0=1.0$ and $A_0=2.0$. In these configurations, two types of elongated small-scale vortices prevail, namely, parallel to the waviness (spanwise vortices) and longitudinally oriented according to the flow direction (streamwise vortices). The former are stemming just behind the tops of the waviness due to a roll-up process induced by the shear force. The size of the structures clearly increases as a function of $A_0$. The fluid, which accelerates on the waviness tops (see the vortices' colour), interacts with a low-speed flow in the valleys. A close inspection of this region reveals the occurrence of small separation bubbles on the downhill sides of the tops in which the streamwise velocity is negative.

\subsection{Effect of the corrugated wall on the flow in the vicinity of separation}

The introduced surface modification affects the flow just above the corrugation as well as after a transition to the smooth surface. The analysis of the flow characteristics will be then carried out in these two regions. First, the area behind the corrugation will be examined and then attention will be paid to changes in the statistical parameters of the boundary layer above the modified surface.

\begin{figure}[ht]
    \centering
    \includegraphics[width=0.6\textwidth]{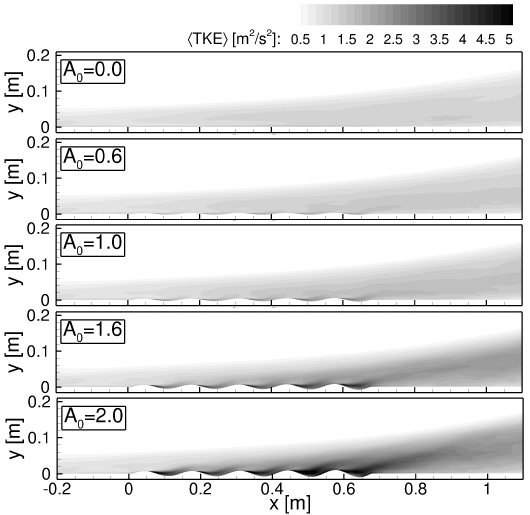}
    \caption{Turbulent kinetic energy maps for $A_0=var.$ and $N_{\lambda}=5$.}
    \label{fig:effect-amplitudes-1}
\end{figure}

Figure \ref{fig:effect-amplitudes-1} presents the contour maps of TKE at the streamwise cross-section in the middle of the domain for five values of $A_0$ and under a constant value of $N_{\lambda}=5$. It can be observed that the growth in $A_0$ causes TKE enhancement which is especially visible for cases with $A_0>1.0$. For these configurations, a considerable increase in TKE originates from waviness crests. This phenomenon agrees with the observations from the DNS of Yoon et al. \cite{yoon2009effect}. The LES research conducted by Zhang et al. \cite{zhang2022numerical,zhang2021large} also showed that increasing the corrugation amplitude leads to the TKE increase in the flow in a similar manner as it is observed in Fig.~\ref{fig:effect-amplitudes-1}. 

\begin{figure}[h!] 
     \begin{subfigure}{0.495\linewidth}
       \includegraphics[trim={0 0 0 0}, clip, width=0.95\linewidth]{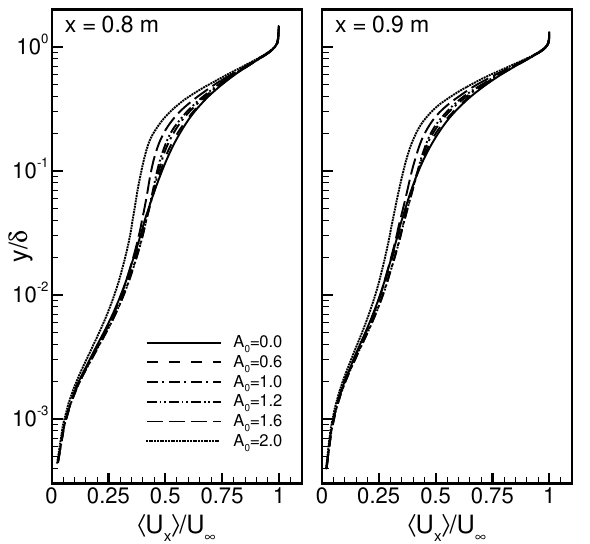}
       \caption{}
       \end{subfigure}
        \begin{subfigure}{0.495\linewidth}
       \includegraphics[trim={0 0 0 0}, clip, width=0.95\linewidth]{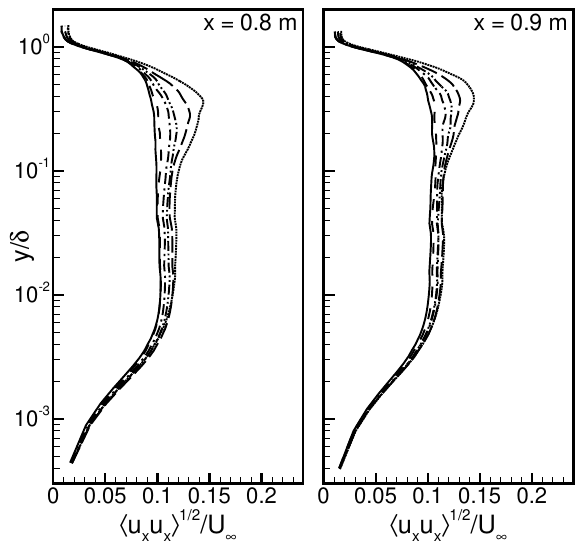}
       \caption{}
      \end{subfigure}
    \caption{Comparison of the mean streamwise velocity (a) and its fluctuations (b) for variable $A_0$ and a constant $N_{\lambda}=5$ at $x=0.8\ m$ and $x=0.9\ m$.}
    \label{fig:effect-downstream-3}
\end{figure}

Figure \ref{fig:effect-downstream-3} presents the evolution of the mean streamwise velocity and its fluctuations with the changing $A_0$ taken for two traverses $x=0.8$ m and $x=0.9$ m. It can be observed that the velocity, normalised by the free-stream velocity, decreases with the amplitude growth for $0.001<y/\delta<0.6$. Figure \ref{fig:effect-downstream-3}b shows that the velocity fluctuations increase with growing $A_0$ and the location of clearly seen outer maximum moves further away from the wall which is associated with the large flow scale activity (see for instance Refs \cite{bobke_vinuesa_örlü_schlatter_2017,NIEGODAJEW2019108456} for more details). These results are consistent with the behaviour and development of vortical structures observed in Fig. \ref{fig:q-param}. 

\begin{figure}[h!] 
     \begin{subfigure}{0.495\linewidth}
       \includegraphics[trim={0 0 0 0}, clip, width=0.95\linewidth]{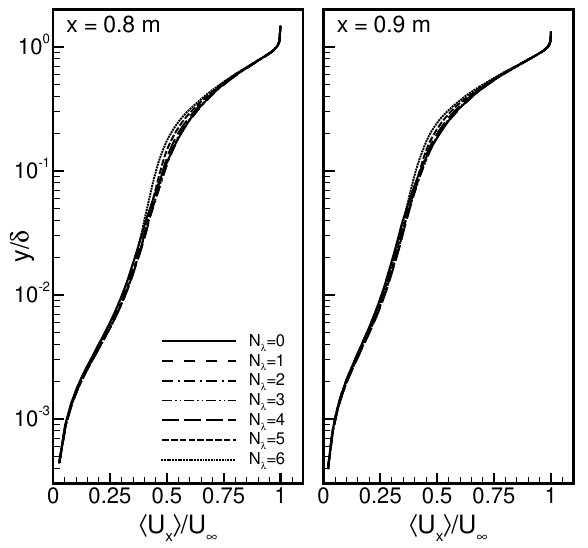}
       \caption{}
       \end{subfigure}
        \begin{subfigure}{0.495\linewidth}
       \includegraphics[trim={0 0 0 0}, clip, width=0.95\linewidth]{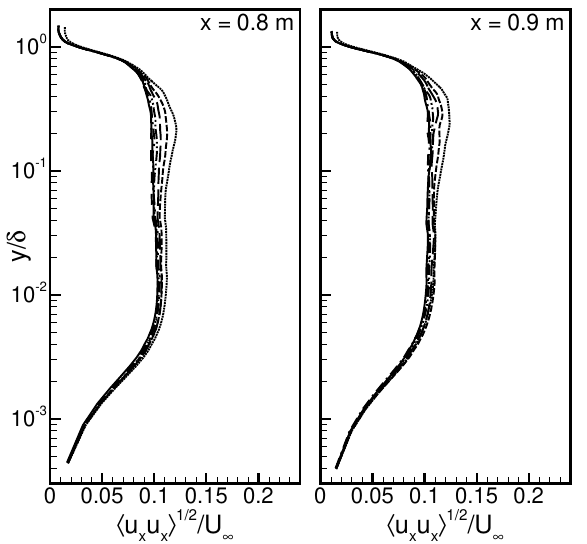}
       \caption{}
      \end{subfigure}
    \caption{Comparison of the mean streamwise velocity (a) and its fluctuations (b) for variable $N_{\lambda}$ and a constant $A_0=1.0$ at $x=0.8\ m$ and $x=0.9\ m$.}
    \label{fig:effect-downstream-4}
\end{figure}

\begin{figure}[h!] 
     \begin{subfigure}{0.9\linewidth}
       \includegraphics[trim={0 0 0 0}, clip, width=\linewidth]{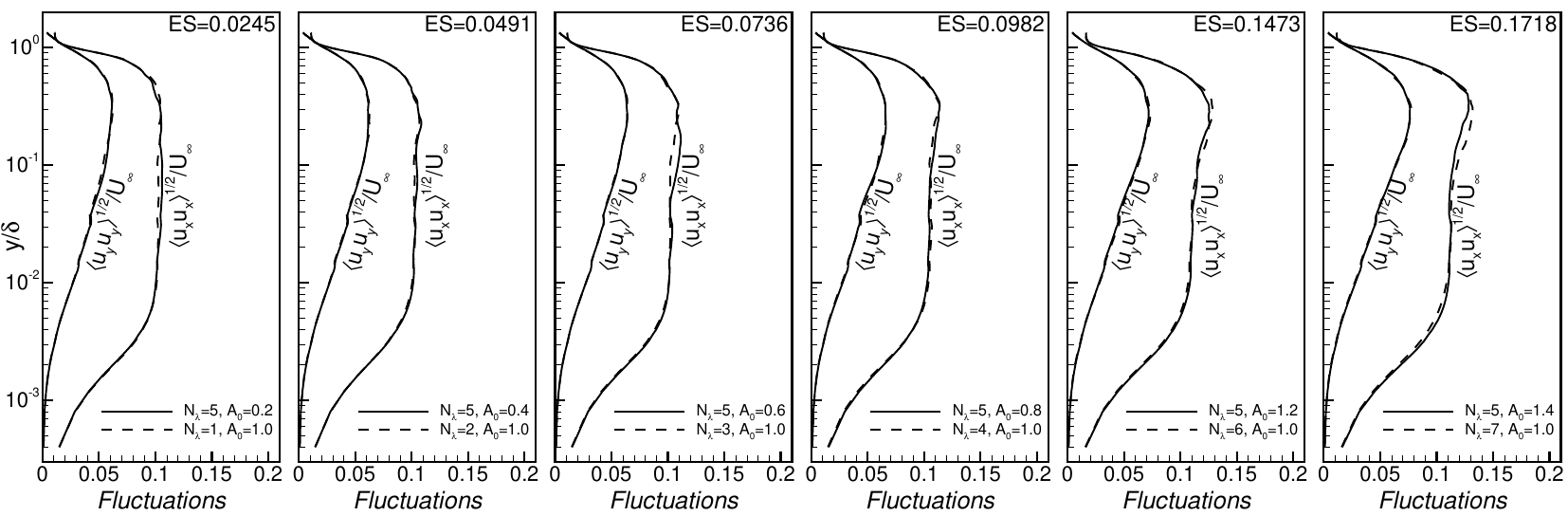}
       \end{subfigure}
    \caption{Evolution of $\langle u_xu_x \rangle ^{1/2}/U_{\infty}$ and $\langle u_yu_y \rangle ^{1/2}/U_{\infty}$ profiles for variable $A_0$ and $N_{\lambda}$ for six different $ES$ values obtained at $x=0.9 $ m.}
    \label{fig:effective-slope-1}
\end{figure}

Figure \ref{fig:effect-downstream-4} presents the comparison of the mean streamwise velocity and its fluctuations for different values of $N_{\lambda}$ and $A_0=1.0$. As $N_{\lambda}$ increases, the velocity at $0.001<y/\delta<0.6$ decreases slightly. When it comes to fluctuations, also an increase is seen with growing $N_{\lambda}$ which is the most pronounced at the location of the outer maximum. The effect of the corrugation on $\langle U_x \rangle$ and $\langle u_xu_x \rangle ^{1/2}$ profiles seems to be the same regardless of whether changing $A_0$ or $N_{\lambda}$. This may suggest that there is another representative factor involving both already mentioned parameters. An interesting candidate for further exploration seems to be $ES$, frequently used in other works to describe surface irregularities \cite{de2016large,forooghi2017toward}. Such an analysis is illustrated in Fig. \ref{fig:effective-slope-1} where it is clearly seen that for cases with different $A_0$ and $N_{\lambda}$ but with the same $ES$, the fluctuation profiles are virtually identical. 

One of the main goals of the investigation was to find such a combination of $A_0$ and $N_{\lambda}$ allowing for the highest increase in $\tau_w$ downstream. In the experiment \cite{drozdz2021effective}, the waviness which ensured the highest wall-shear increase was found to be characterised by $A_0=1.0$ and $N_{\lambda}=5$ for $Re_{\tau}=4000$. Although in the present study a lower value of $Re_{\tau}=2500$ is studied, the above-mentioned values of $A_0$ and $N_{\lambda}$ were used as reference parameters. 

\begin{figure}[h!] 
    \centering
    \begin{subfigure}{0.49\linewidth}
    \includegraphics[trim={0 0 0 0}, clip, width=\linewidth]{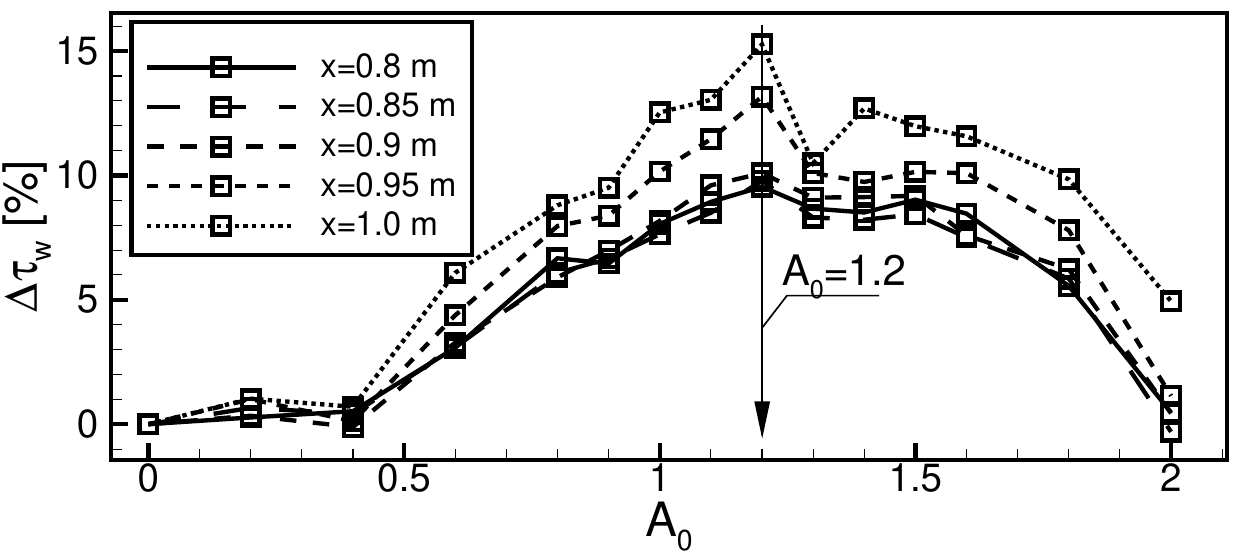}
    \caption{}
    \end{subfigure}
    \begin{subfigure}{0.49\linewidth}
    \includegraphics[trim={0 0 0 0}, clip, width=\linewidth]{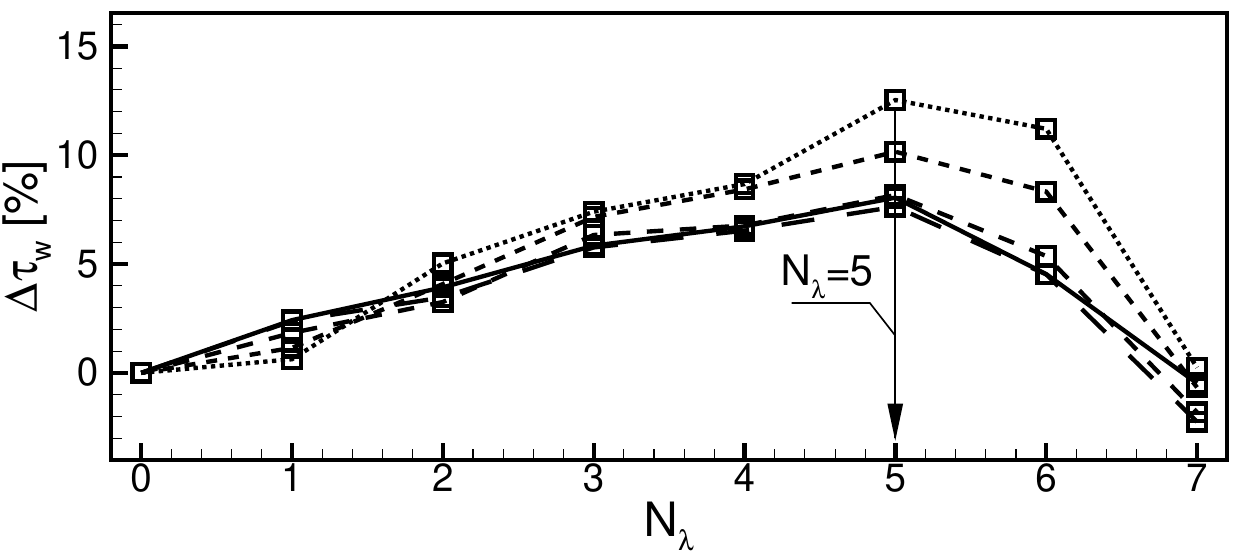}
    \caption{}
    \end{subfigure}
    \caption{Wall-shear stress increase for variable $A_0$ (a) and $N_{\lambda}$ (b) at different distances from $x_0$.}
    \label{fig:effect-downstream-1}
\end{figure}

Figure \ref{fig:effect-downstream-1}a presents the effect of $A_0$ under constant $N_{\lambda}=5$ on the wall-shear stress for several selected streamwise distances, i.e. $x=0.8$ m, $0.85$ m, $0.9$ m, $0.95$ m, $1.0$ m. It can be observed that with the increase in $A_0$, $\Delta\tau_w$ also increases until $A_0=1.2$. However, with further growth in $A_0$, $\Delta\tau_w$ starts decreasing. A local minimum is observed at $A_0=1.3$ for all of the traverses which, however, becomes less and less visible with decreasing distance from the waviness. It indicates that compared to the case with $A=0$, the surface waviness with $A_0=1.2$ corresponding to $A^+=139$ (Tab. \ref{Tab:Adiff}), yields the highest enhancement of $\Delta\tau_w$ ranging from $9\%$ up to $15.5\%$, depending on the traverse ($x$ location). Figure \ref{fig:effect-downstream-1}b presents, on the other hand, the effect of $N_{\lambda}$ when keeping constant $A_0=1.0$ on $\tau_w$ for various $x-$distances. It can be observed that the most beneficial effect appears for $N_{\lambda}=5$. For $N_{\lambda}>5$, the decrease in $\Delta \tau_w$ can be noted. 

\newpage
\begin{figure}[h!] 
    \centering
    \begin{subfigure}{0.65\linewidth}
    \includegraphics[trim={0 0 0 0}, clip, width=\linewidth]{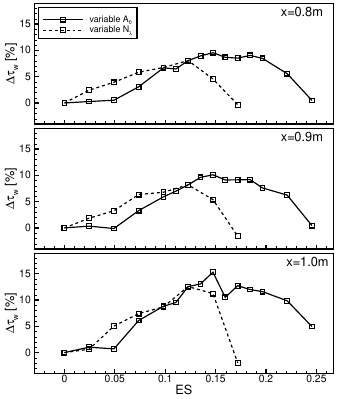}
    \end{subfigure}
    \caption{Wall-shear stress increase for variable $A_0$ (a) and $N_{\lambda}$ (b) at different distances from $x_0$.}
    \label{fig:effective-slope-2}
\end{figure}

The results presented so far may suggest that $ES$ is the most representative parameter (rather than $A_0$ or $N_{\lambda}$) that determines the changes in the near-wall flow since the collapse of velocity fluctuation profiles is seen for fixed $ES$ and different combinations of $A_0$ or $N_{\lambda}$ in \fig{fig:effective-slope-1}. It is therefore worth checking whether there is a correlation between $ES$ and $\Delta\tau_w$. To verify it, Fig. \ref{fig:effective-slope-2} shows the already presented data (Fig. \ref{fig:effect-downstream-1}) but this time as a function of $ES$, for three selected traverses, i.e. $x=0.8$ m, $0.9$ m and $1.0$ m. The results indicate that there is an increase in $\Delta\tau_w$ up to $ES \approx 0.12$ regardless of whether changing either $A_0$ or $N_{\lambda}$. However, for $ES>0.12$ $\Delta\tau_w$ starts decreasing when changing $N_{\lambda}$. This might indicate that as long as the near-wall flow is dominated by viscous drag (waviness regime \cite{napoli2008effect,schultz2009turbulent,nugroho_monty_utama_ganapathisubramani_hutchins_2021}), the enhancement of $\tau_w$ may be expected.

\begin{figure}[h!] 
    \centering
    \begin{subfigure}{0.5\linewidth}
    \includegraphics[trim={0 0 0 0}, clip, width=\linewidth]{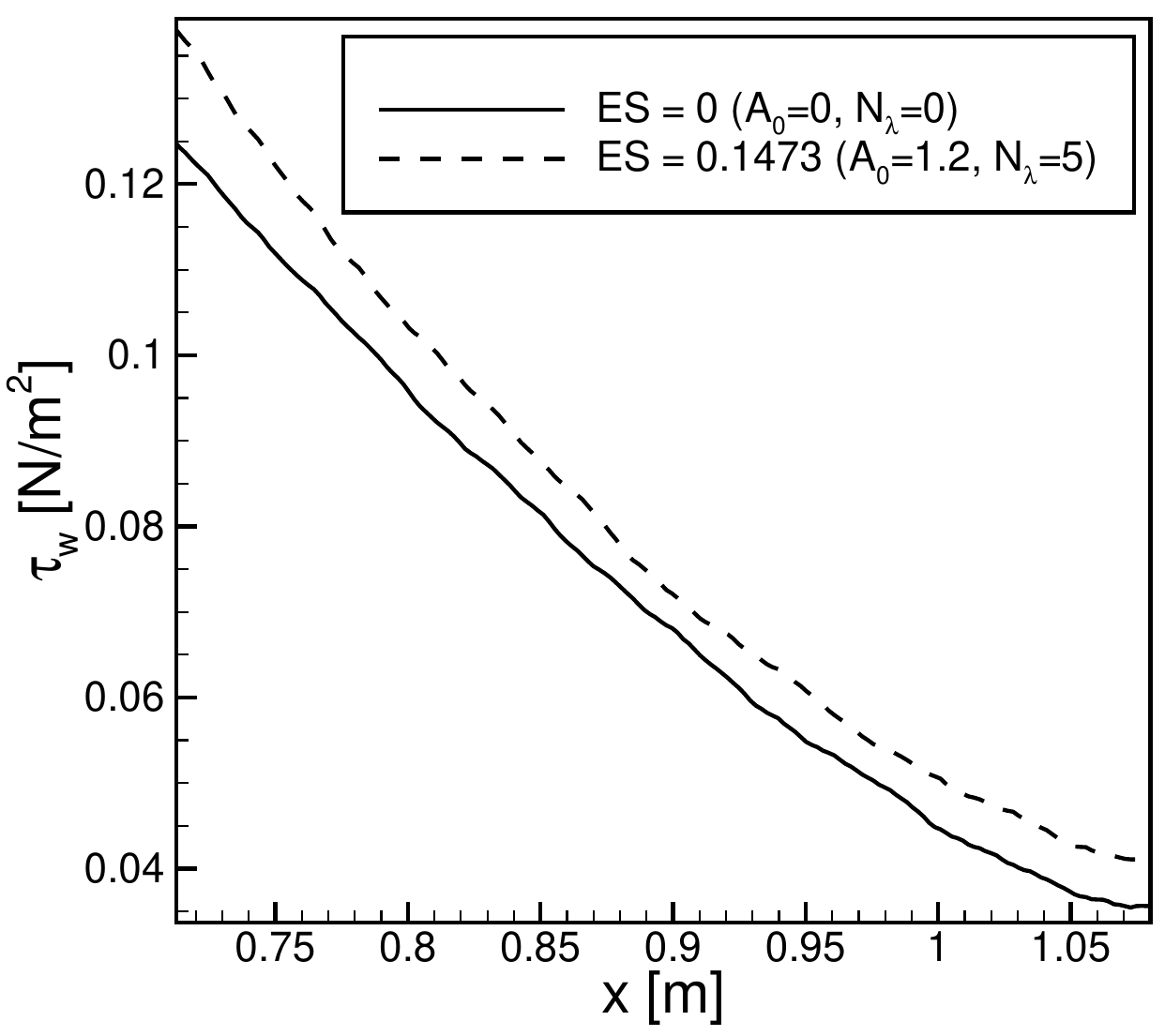}
    \end{subfigure}
    \caption{Wall-shear stress distribution for the optimal $ES=0.1473$ ($N_{\lambda}=5$, $A_0=1.2$) with respect to the flat wall configuration ($A_0=0$).}
    \label{fig:effect-downstream-5}
\end{figure}

As indicated by Dr{\' o}{\. z}d{\. z} et al. \cite{drozdz2021effective}, the optimal corrugation for $Re_{\tau} = 4000$ was $A_0=1.0$ and $N_{\lambda}=5$ ($ES=0.1227$). For higher $ES$ values, the beneficial effect quickly disappears. In the present numerical investigation, for $Re_{\tau}=2500$, the corrugation which provides the highest increase in wall-shear stress is $A_0=1.2$ and $N_{\lambda}=5$ ($ES=0.1473$). The distribution of $\tau_w$ for this case is presented in Fig. \ref{fig:effect-downstream-5}. An increment in $\tau_w$ caused by the corrugation is kept relatively constant throughout the flat part of the domain downstream of the waviness.

\begin{table}[ht]
\begin{tabular}{p{140pt}p{100pt}p{70pt}p{40pt}p{40pt}p{40pt}}
\hline
\hline
$Source$ & $Type$ & $U_{\infty}\ [m/s]$ & $Re_{\tau}$ & $ES$ & $A^+$ \\ \hline
Elsner et al. (2022) \cite{elsner2022experimental} & Experiment/LES & 6 & 1350 & -------- & -------- \\ 
Present study (2023) & LES & 15 & 2500 & 0.1473 & 139\\ 
Dr{\' o}{\. z}d{\. z} et al. (2018) \cite{drozdz2018passive} & Experiment & 20 & 3300 & 0.1242 & 151 \\ 
Dr{\' o}{\. z}d{\. z} et al. (2021) \cite{drozdz2021effective} & Experiment & 24 & 4000 & 0.1227 & 170\\ 
\hline\hline
\end{tabular}
\caption{Parameters of wavy wall ensuring the highest growth in $\tau_w$ for previous databases and the present study.}
\label{Tab:effective-slope-retau}
\end{table}

The obtained results reported in \cite{de2016large,forooghi2017toward}, indicate that a two-dimensional corrugated surface is effective in preventing turbulent separation although its effectiveness depends on $Re$ and $ES$. 
In Table \ref{Tab:effective-slope-retau}, it can be observed that $ES$ which provides the highest increase in $\tau_w$ compared to the flat wall configuration decreases with growing $Re_{\tau}$. The opposite trend is observed with $A^+$, which increases with $Re$. Interestingly, it seems that the corrugations ensuring the highest growth in the wall-shear stress are characterised by amplitudes as high as the size of the near-wall region, namely, the part of TBL up to the logarithmic region, i.e. up to $y^+=150$ \cite{buschmann2006recent}. So, $ES$ that maximises the wall-shear stress decreases with $Re$, which is consistent with the observations of Nugroho et al \cite{nugroho_monty_utama_ganapathisubramani_hutchins_2021}. Moreover, in the work \cite{elsner2022experimental}, neither of the corrugations studied for a wide range of $A_0$ ensured an increase in $\tau_w$ for $Re_{\tau}=1350$. The above discussion suggests that below a certain value of $Re_{\tau}$ the wavy wall becomes ineffective in enhancing the wall-shear stress in the presence of a strong APG. This can be explained by referring to the work of Dr{\' o}{\. z}d{\. z} et al. \cite{drozdz2021effective} who showed a diminishing effect of the so-called amplitude modulation on APG TBL (see Ref. \cite{mathis2009comparison} for more details), in particular on the near-wall flow, with decreasing $Re$. 

\subsection{Insight into the flow above the corrugation}

\begin{figure}[h!] 
    \centering
    \begin{subfigure}{0.75\linewidth}
    \includegraphics[angle=90,origin=c, trim={0 0 0 0}, clip, width=\linewidth]{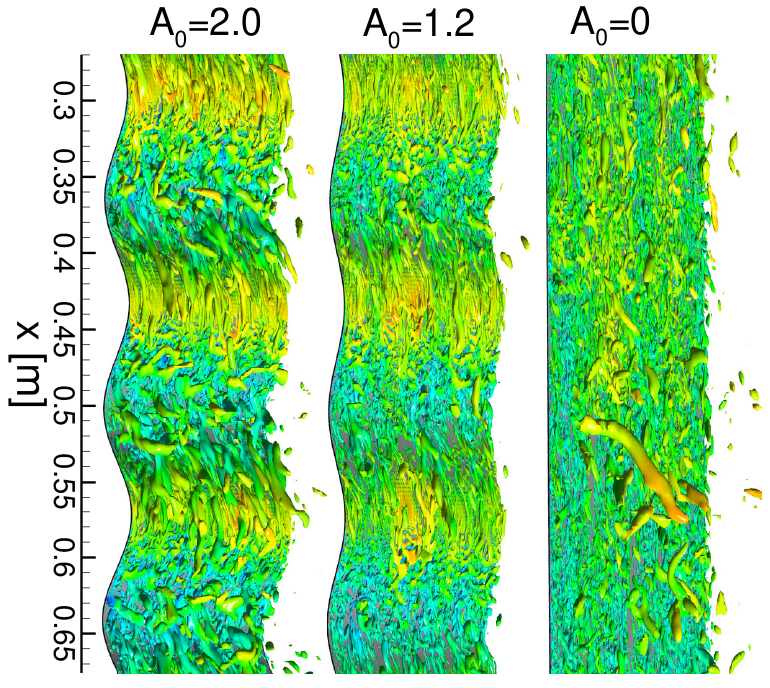}
    \end{subfigure}
\caption{Q--parameter isosurface ($Q=0.05$~s$^{-2}$) coloured by the instantaneous streamwise velocity for cases $A_0=0$, $A_0=1.2$, $A_0=2.0$ at a constant $N_{\lambda}=5$.}
\label{fig:q-param-2}
\end{figure}

The previous section 3.1 was focused on the effect of the corrugation parameters ($A_0$, $N_{\lambda}$ and $ES$) on the flow downstream the modified surface, i.e. in the streamwise region from the end of the waviness towards the outlet of the computational domain (from $x=0.676$ m to $x=1.1$ m). This section summarises how these parameters affect the flow and most importantly the wall-shear stress on the waviness location (from $x=0.01$ m to $x=0.676$ m). The analysis is focused on the point of the separation within subsequent corrugation periods. No less interesting is also the separation bubble length inside each trough as well as how $A_0$ and $N_{\lambda}$ contribute to detachment point locations. 

Figure \ref{fig:q-param-2} shows the zoom of the $Q-$parameter isosurface for three selected cases, ranging from $x=0.27$ m to $x=0.676$ m. For the flat plate ($A_0=0$), small ordered vortices with fragments of hairpin-type structures can be identified at $x\approx 0.3$ m. Further downstream, as APG becomes stronger, the near-wall turbulent activity decreases and larger structures appear in the outer region of the boundary layer. For cases with $A_0>0$, a lot of longitudinal structures appear on the uphill side of the crest, which is consistent with observation from \cite{de1997direct}. Also downhill side of the crest generates spanwise vortices. Further downstream, for $A_0=1.2$ and $A_0=2.0$, an increase in the size of structures in the outer region of the boundary layer can also be seen. A close inspection of this region reveals the occurrence of small separation bubbles on the downhill sides of the waviness in which the streamwise velocity is negative. The activity of the spanwise vortices is strong downhill (local APG effect) which is consistent with previous observations \cite{yoon2009effect, zhang2021effects, zhang2022numerical}. The stronger activity of the spanwise vortices in APG prevents the flow separation (for moderate $ES$) as the sweep events generated by these spanwise vortices are stronger which is the result of a stronger convection velocity \cite{drozdz2023convection}. The second type of vortices exists on the uphill side, where the local favourable pressure gradient (FPG) occurs. These are the quasi-streamwise type structures, similar to ones that can be observed for canonical TBLs \cite{SCHLATTER201475,schoppa_hussain_2002} at high Reynolds number, which are generated downstream the reattachment point (documented later) in the local FPG region generated by waviness. Furthermore, as observed by De Angelis et al. \cite{de1997direct}, when $A/\lambda$ increases, the reattachment point moves towards the crest ahead and so, the region occupied by the quasi-streamwise vortices becomes reduced.

\begin{figure}[h!] 
    \centering
    \begin{subfigure}{0.7\linewidth}
    \includegraphics[trim={0 0 0 0}, clip, width=\linewidth]{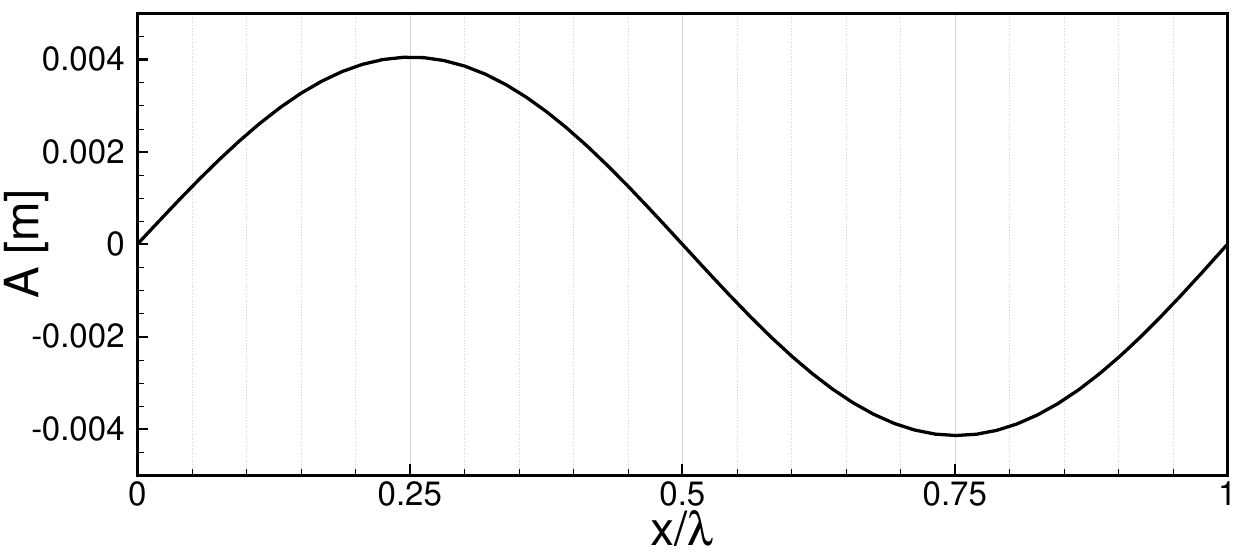}
    \caption{}
    \end{subfigure}
    \caption{Position of the waveform as a function of $x_{sep}/\lambda$.}
    \label{fig:amplitude-schematic}
\end{figure}

For convenience, let us introduce the non-dimensional parameter $x/\lambda$ that is used further to explore the results above the waviness. The graphical interpretation of this parameter is presented in Fig.\ref{fig:amplitude-schematic}. 

\begin{figure}[h!] 
    \centering
    \begin{subfigure}{0.99\linewidth}
    \includegraphics[trim={0 0 0 0}, clip, width=\linewidth]{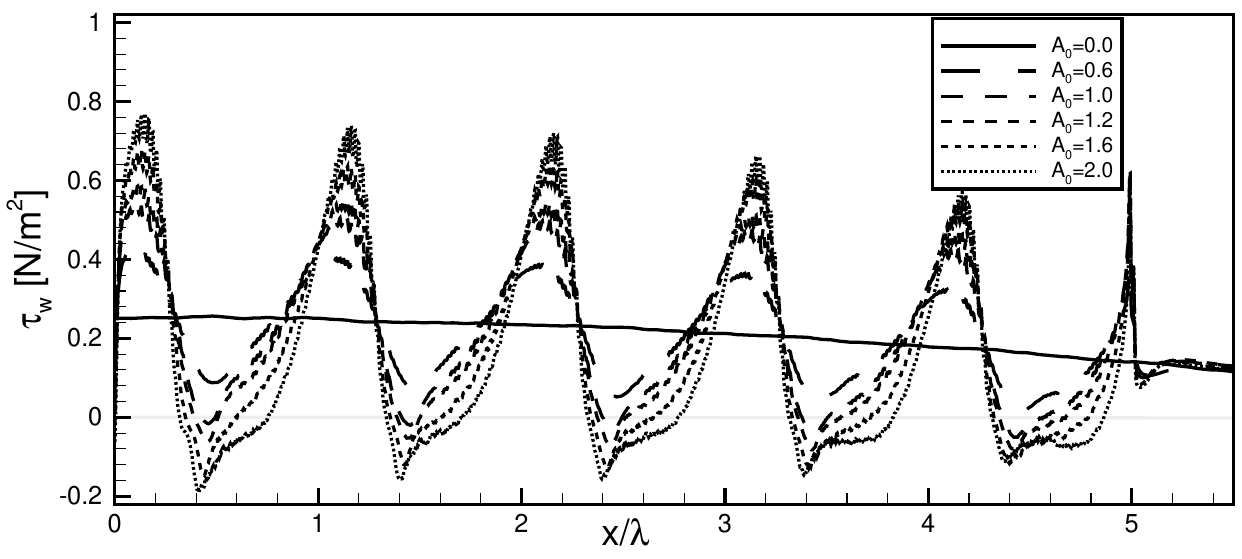}
    \caption{}
    \end{subfigure}
        \begin{subfigure}{0.99\linewidth}
    \includegraphics[trim={0 0 0 0}, clip, width=\linewidth]{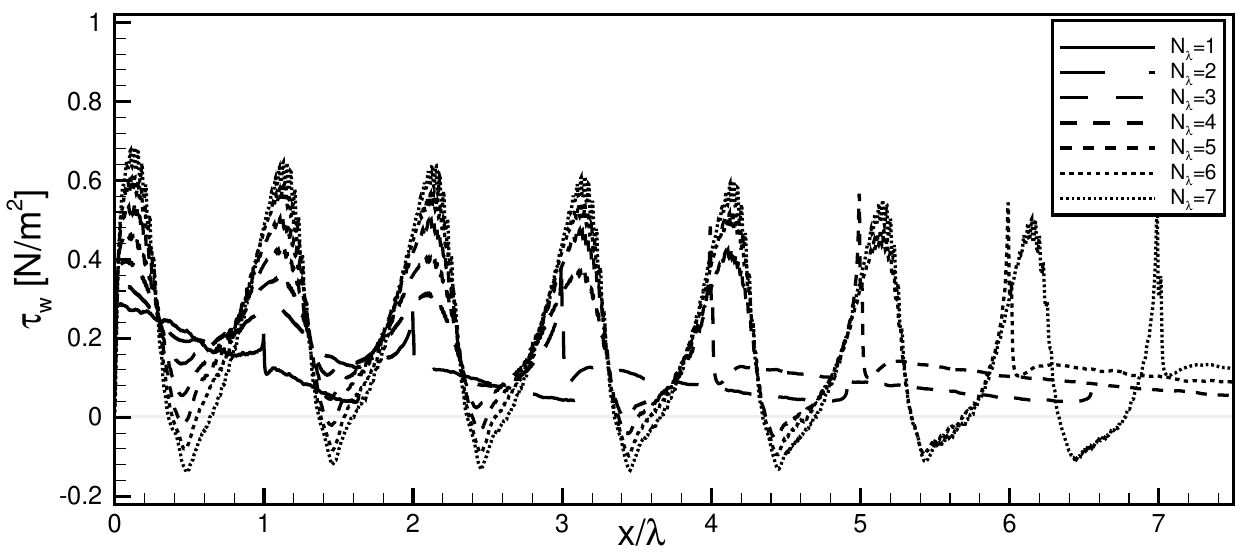}
    \caption{}
    \end{subfigure}
    \caption{Distribution of the $x-$component of wall-shear stress on the wavy wall for variable $A_0$ at constant $\lambda=0.133\ m$ (a) and variable $N_{\lambda}$ at constant $A_0=1.0$ (b). The $x-$axis is the streamwise coordinate normalised by the length of the corrugation period ${\lambda}$ (Tab. \ref{Tab:Adiff},\ref{Tab:NLdiff}.). Gray line indicates $\tau_w=0$.}
    \label{fig:effect-amplitudes-2}
\end{figure}

\begin{figure}[h!] 
    \centering
    \begin{subfigure}{0.4\linewidth}
    \includegraphics[trim={0 0 0 0}, clip, width=\linewidth]{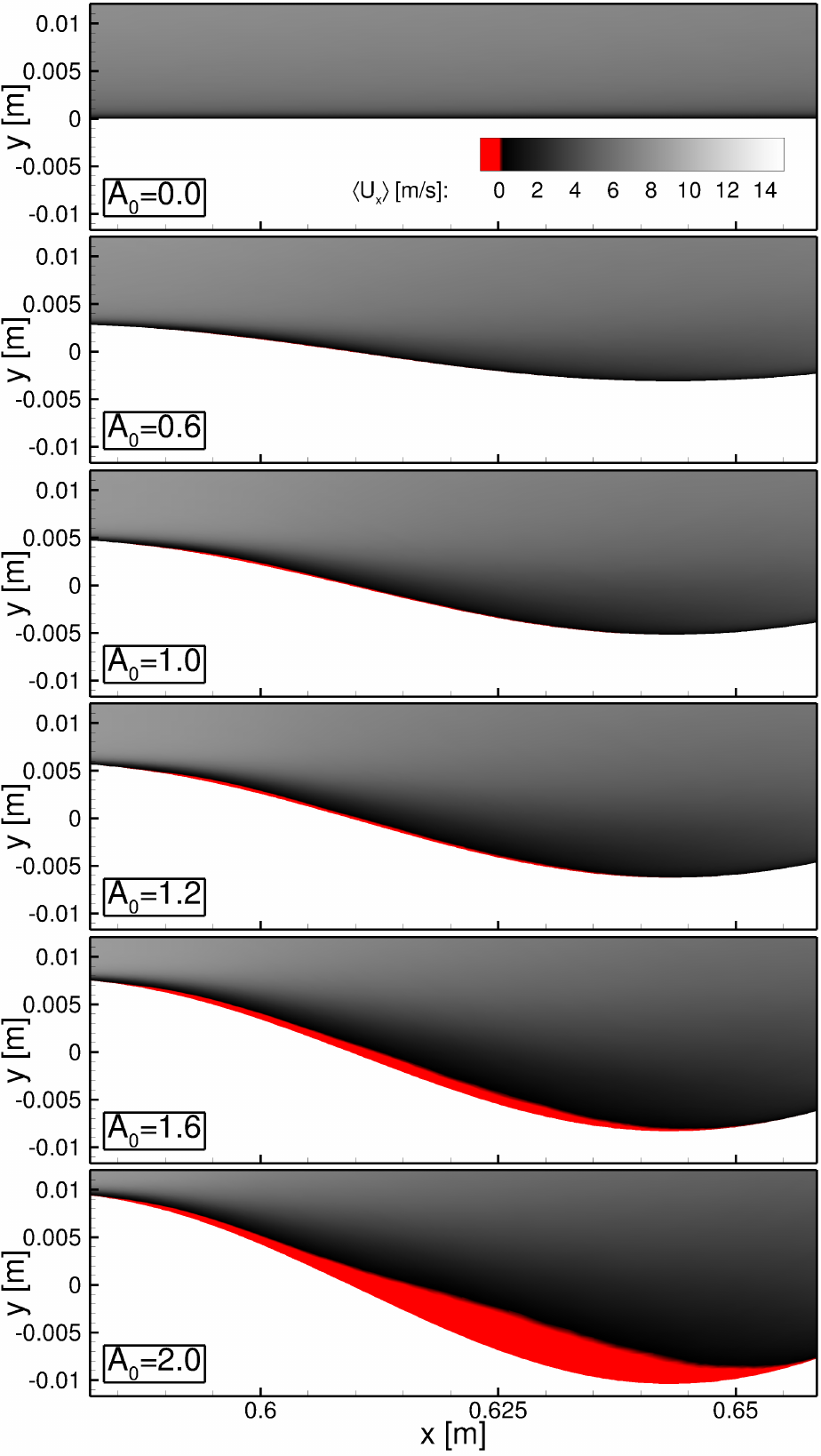}
    \end{subfigure}
    \caption{Contour maps of $\langle U_x \rangle$ at the last trough of the corrugation for the chosen $A_0$ cases at a constant $N_{\lambda}=5$. The values of $\langle U_x \rangle < 0$ m/s, indicating the separation bubble, are marked with red colour.}
    \label{fig:effect-amplitudes-8}
\end{figure}

Figure \ref{fig:effect-amplitudes-2} presents the distribution of the $x-$component of the wall-shear stress along the corrugated surface for different amplitudes (a) and different numbers of periods (b). The streamwise distance $x$ is normalised by the length of the waviness period $x/\lambda$. The bold line indicates the distribution of $\tau_w$ for the reference case $A_0=0$ (flat plate) while the grey horizontal line indicates $\tau_w=0$. The streamwise decrease in $\tau_w$ for the flat plate case is due to the presence of APG. When corrugation emerges, even for small values of $A_0$ and $N_{\lambda}$ and thus for low $ES$, a streamwise oscillation of the wall-shear stress is seen. The maximal values of $\tau_w$ are just before the waviness crests, while the minimal shear is noted just before troughs. An increase either in $A_0$ or $N_{\lambda}$ leads to a strong reaction of the flow manifested by enhanced variability of $\tau_w$. In Fig. \ref{fig:effect-amplitudes-2}a, for $A_0>1.0$, the separation is seen in all troughs (manifested by $\tau_w<0$). The separation is less likely to occur when changing $N_{\lambda}$ (Fig. \ref{fig:effect-amplitudes-2}b), which may be attributed to the fact that the analysed $ES$ range is smaller than in Fig. \ref{fig:effect-amplitudes-2}a (see Table~\ref{Tab:Adiff} and Table~\ref{Tab:NLdiff}). In general, an increase in the amplitude of the wall-shear stress oscillation with increasing $A_0$ might indicate that the higher the amplitude of the waviness the stronger the local FPG and local APG. The presence of the local APG causes the flow detachment to occur within the corrugation when $A_0$ and $N_{\lambda}$ are large enough. It can be assumed that the presence of local separation in waviness troughs has a significant effect on the change in $\tau_w$ downstream of the corrugation. To study this effect, the contour maps of the $x-$component of velocity for selected $A_0$ within the last (fifth) trough are presented in Fig. \ref{fig:effect-amplitudes-8}. For convenience of analysis, the area of separation has been highlighted in red. 

\newpage

\begin{figure}[h!] 
    \centering
    \begin{subfigure}{0.48\linewidth}
    \includegraphics[trim={0 0 0 0}, clip, width=\linewidth]{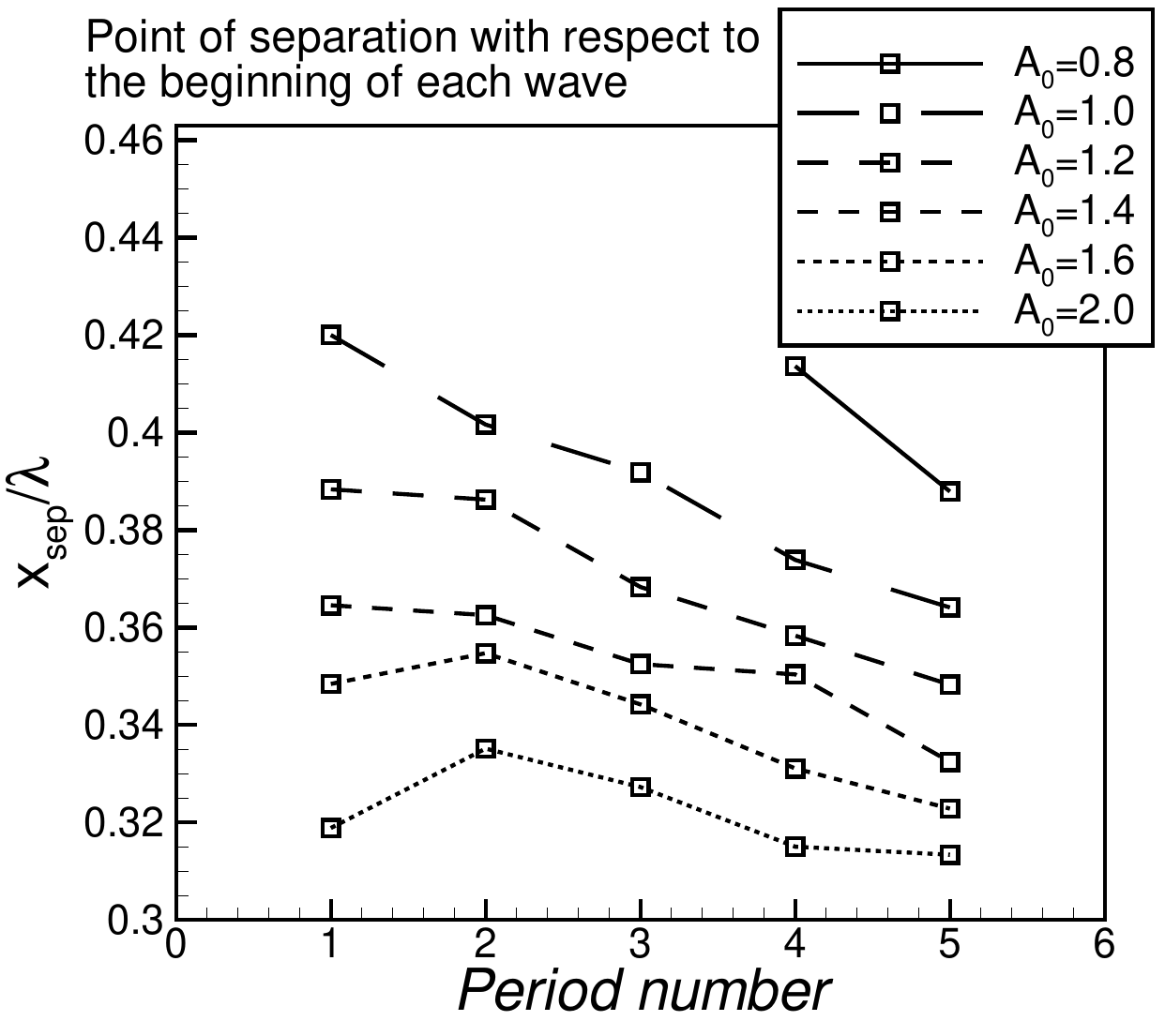}
    \caption{}
    \end{subfigure}
    \begin{subfigure}{0.48\linewidth}
    \includegraphics[trim={0 0 0 0}, clip, width=\linewidth]{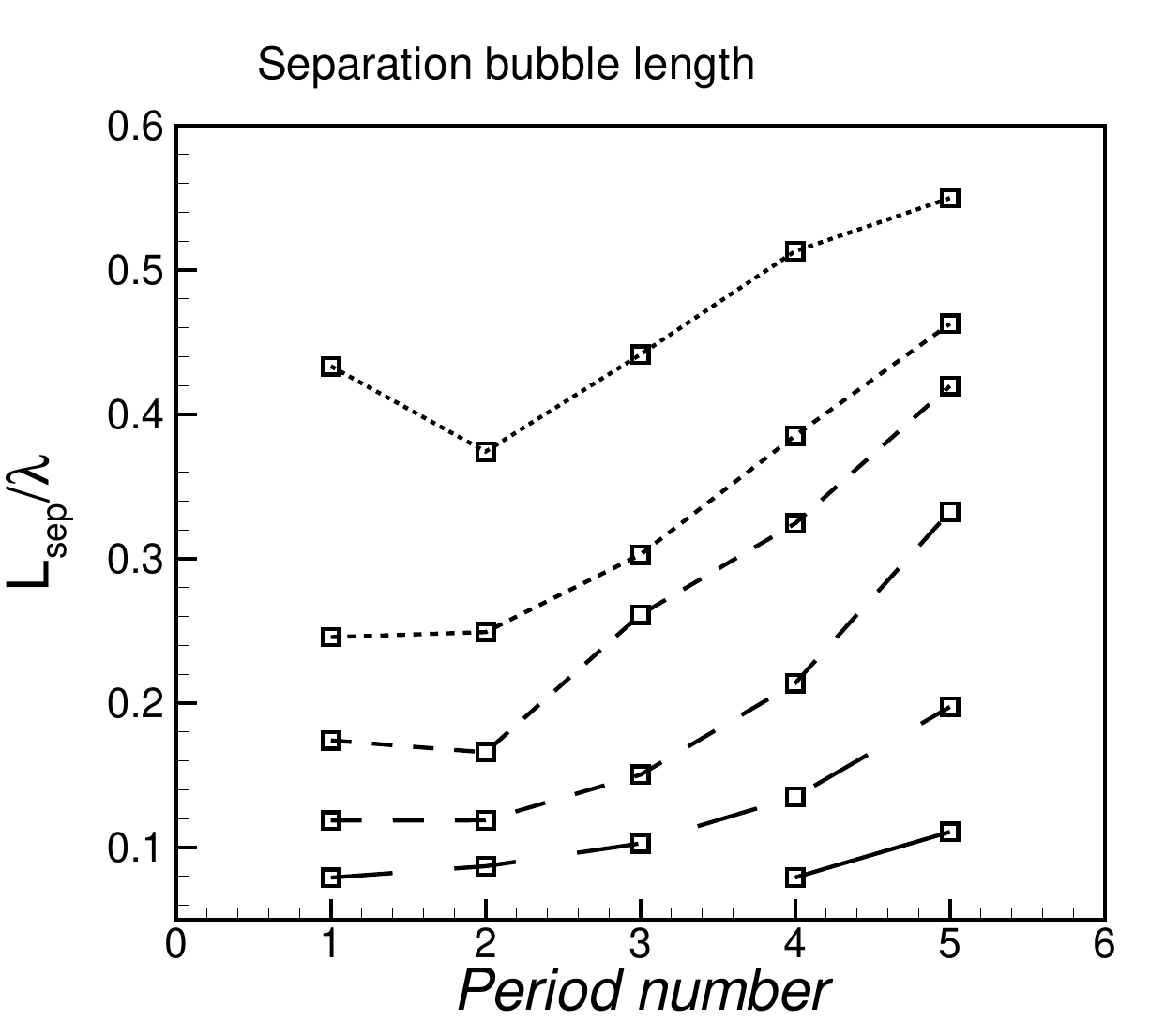}
    \caption{}
    \end{subfigure}
    \caption{The location of the point of separation (a) and the separation bubble length (b) within each wave for variable $A_0$ cases at constant $N_{\lambda}=5$}.
    \label{fig:effect-amplitudes-4}
\end{figure}

Figure \ref{fig:effect-amplitudes-4} presents the location of the point of separation $x_{sep}/\lambda$ (a) and the separation bubble length $L_{sep}/\lambda$ (b) within each subsequent wave. To have a better view of the localisation where the separation occurs within a waviness, please refer to \fig{fig:amplitude-schematic}. It can be observed (in Fig. \ref{fig:effect-amplitudes-4}) that, apart from $A_0=2.0$ case, the point of separation within each consecutive corrugation period moves upstream (towards the crest) with each subsequent wave. This is manifested by decreasing values of $x_{sep}/\lambda$ approaching $x/\lambda=0.25$ (Fig. \ref{fig:amplitude-schematic}) i.e., the corrugation peak. The same effect occurs with growth of $A_0$ and $N_{\lambda}$. This generally means that the separation within subsequent waves occurs earlier because the amplitude of the wavy wall increases downstream. As the detachment occurs earlier, the point of reattachment moves downstream causing the growth of the size of the separation bubble. This is even better illustrated in Fig. \ref{fig:effect-amplitudes-4}b, where one can note that when $A_0$ and $N_{\lambda}$ increase, the separation bubble increases as well. This effect is evidently seen for the higher values of $A_0$. The separation did not occur for amplitudes lower than $A_0=0.8$, so the results for $A_0<0.8$ are not shown here.

\begin{figure}[h!] 
    \centering
    \begin{subfigure}{0.48\linewidth}
    \includegraphics[trim={0 0 0 0}, clip, width=\linewidth]{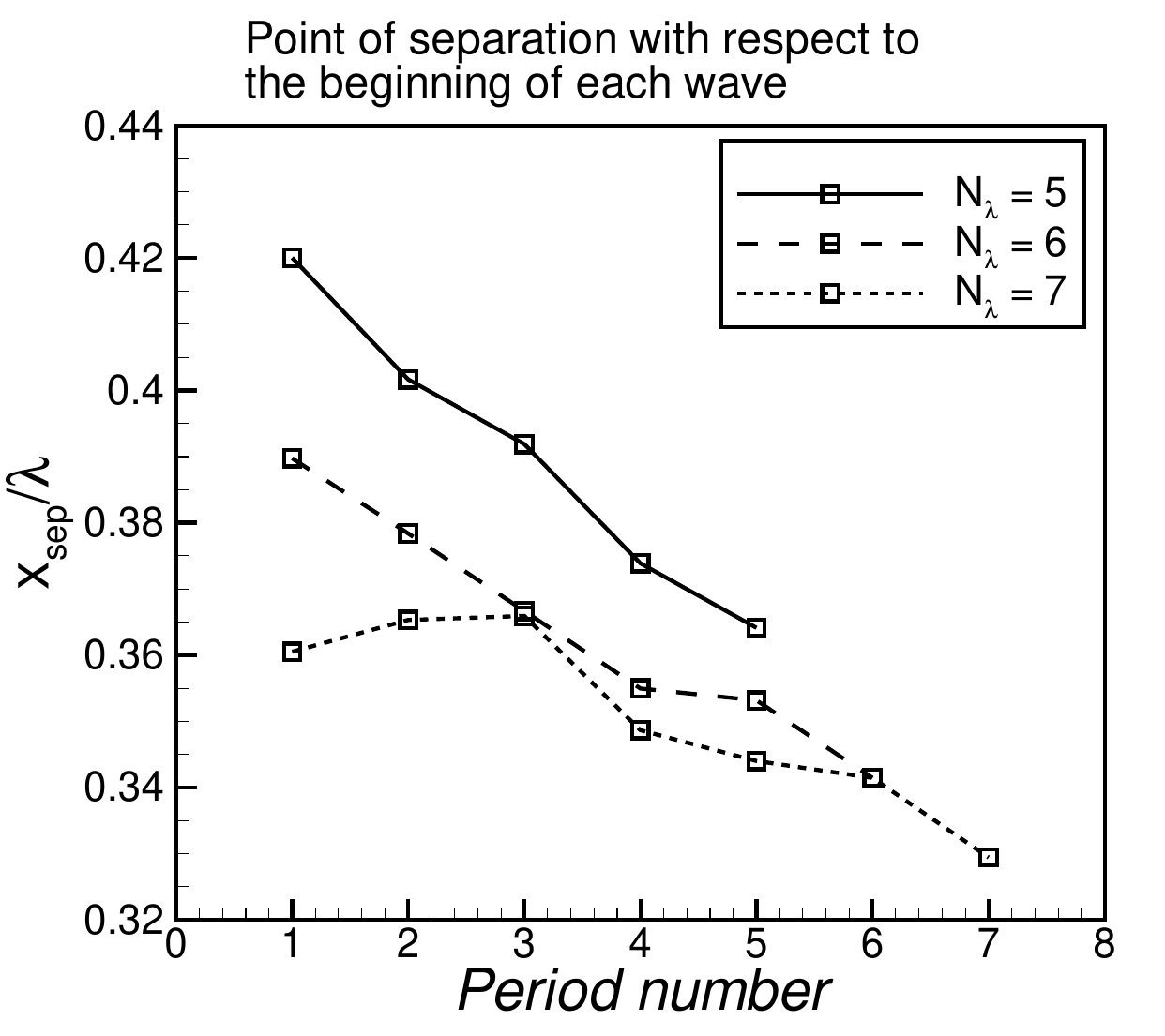}
    \caption{}
    \end{subfigure}
    \begin{subfigure}{0.48\linewidth}
    \includegraphics[trim={0 0 0 0}, clip, width=\linewidth]{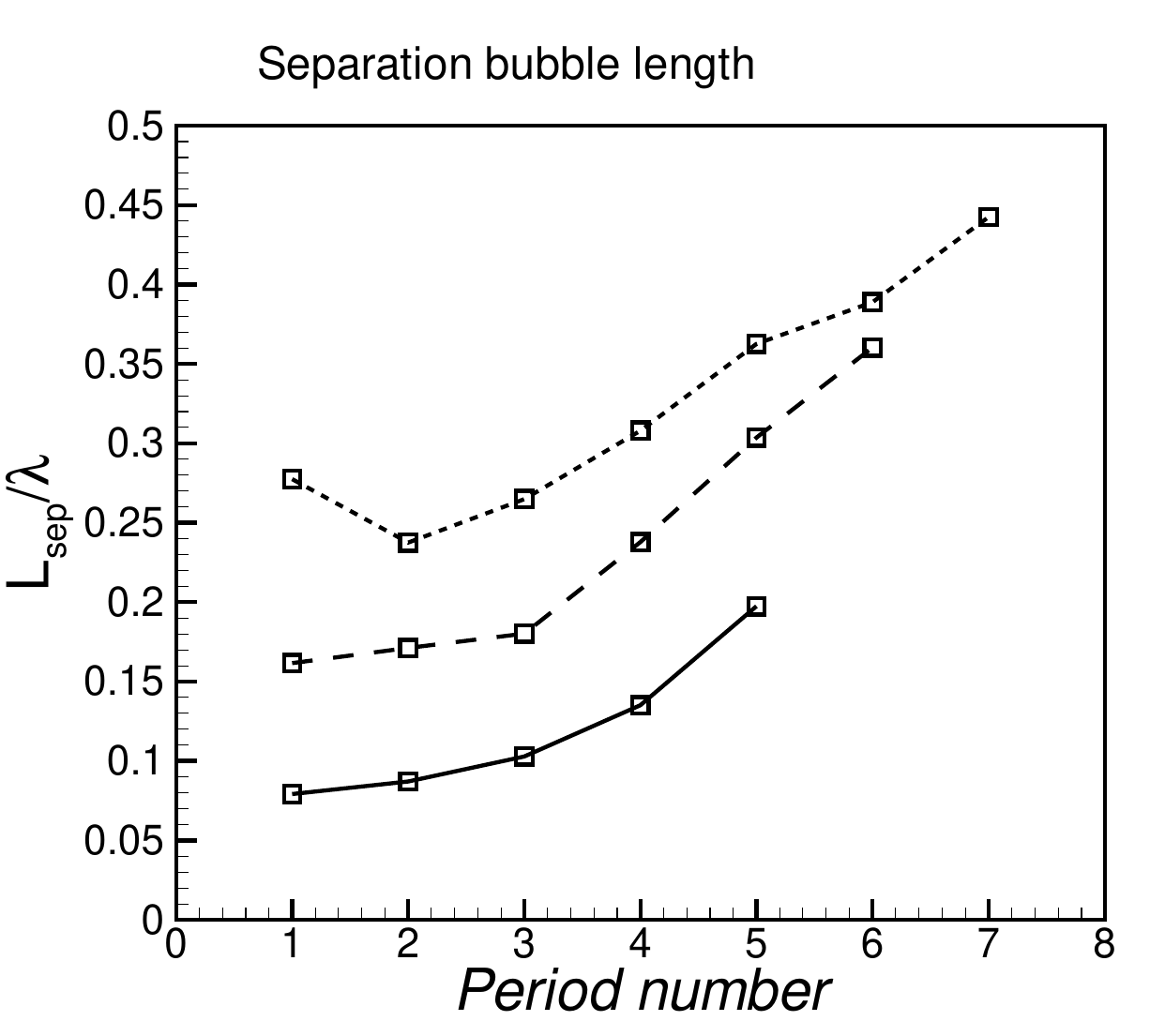}
    \caption{}
    \end{subfigure}
    \caption{The location of the point of separation (a) and the separation bubble length (b) within each wave for variable $N_{\lambda}$ cases at constant $A_{0}=1.0$}.
    \label{fig:effect-amplitudes-7}
\end{figure}

Figure \ref{fig:effect-amplitudes-7} presents the location of the point of separation (a) and the separation bubble length (b) within subsequent waves. It can be observed that the point of flow detachment moves downstream within subsequent periods (decreasing values of $x_{sep}/\lambda$, similarly as with increasing $A_0$ in Fig. \ref{fig:effect-amplitudes-4}) and the separation bubble grows up. It is worth noting, that the flow separation appears when $N_{\lambda}\geq 5$, while for $N_{\lambda}<5$ the flow field is free of that effect. Interestingly, although the case $N_{\lambda}=4$ and $A_0=1.0$ (Table \ref{Tab:NLdiff}) has the same effective slope $ES=0.0982$ and amplitude to period ratio $A/{\lambda}=0.0204$ as the case $A_0=0.8$ and $N_{\lambda}=5$ (Table \ref{Tab:Adiff}) the former one does not lead to separation in any trough. In the latter case, the separation is seen in 4$^{th}$ and 5$^{th}$ trough (see Fig. \ref{fig:effect-amplitudes-4}). This may suggest that neither $ES$ nor $A/\lambda$ can serve as a representative indicator on whether the separation will or will not occur in corrugation troughs.

\begin{figure}[h!] 
    \centering
    \begin{subfigure}{0.48\linewidth}
    \includegraphics[trim={0 0 0 0}, clip, width=\linewidth]{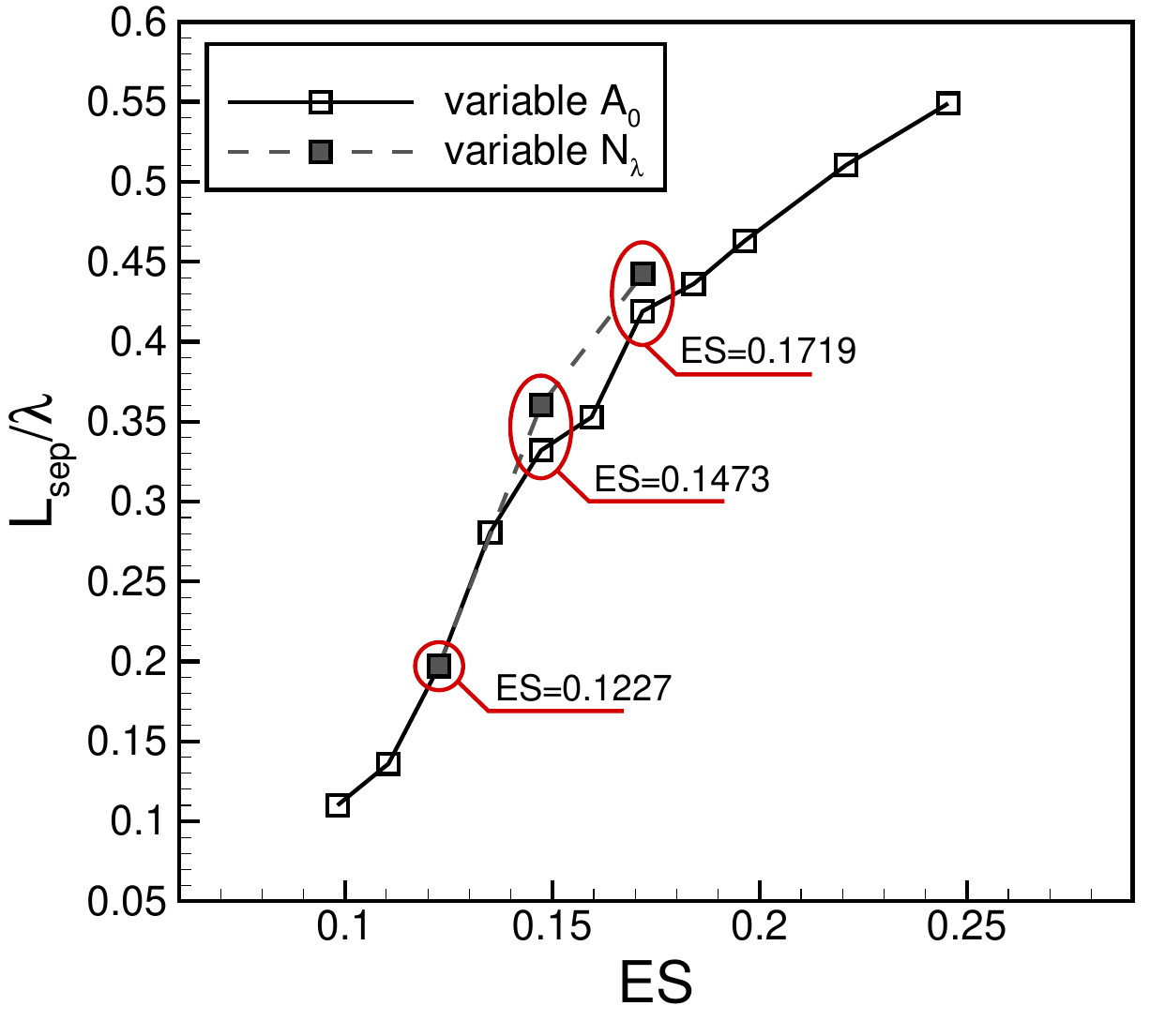}
    \caption{}
    \end{subfigure}
    \begin{subfigure}{0.48\linewidth}
    \includegraphics[trim={0 0 0 0}, clip, width=\linewidth]{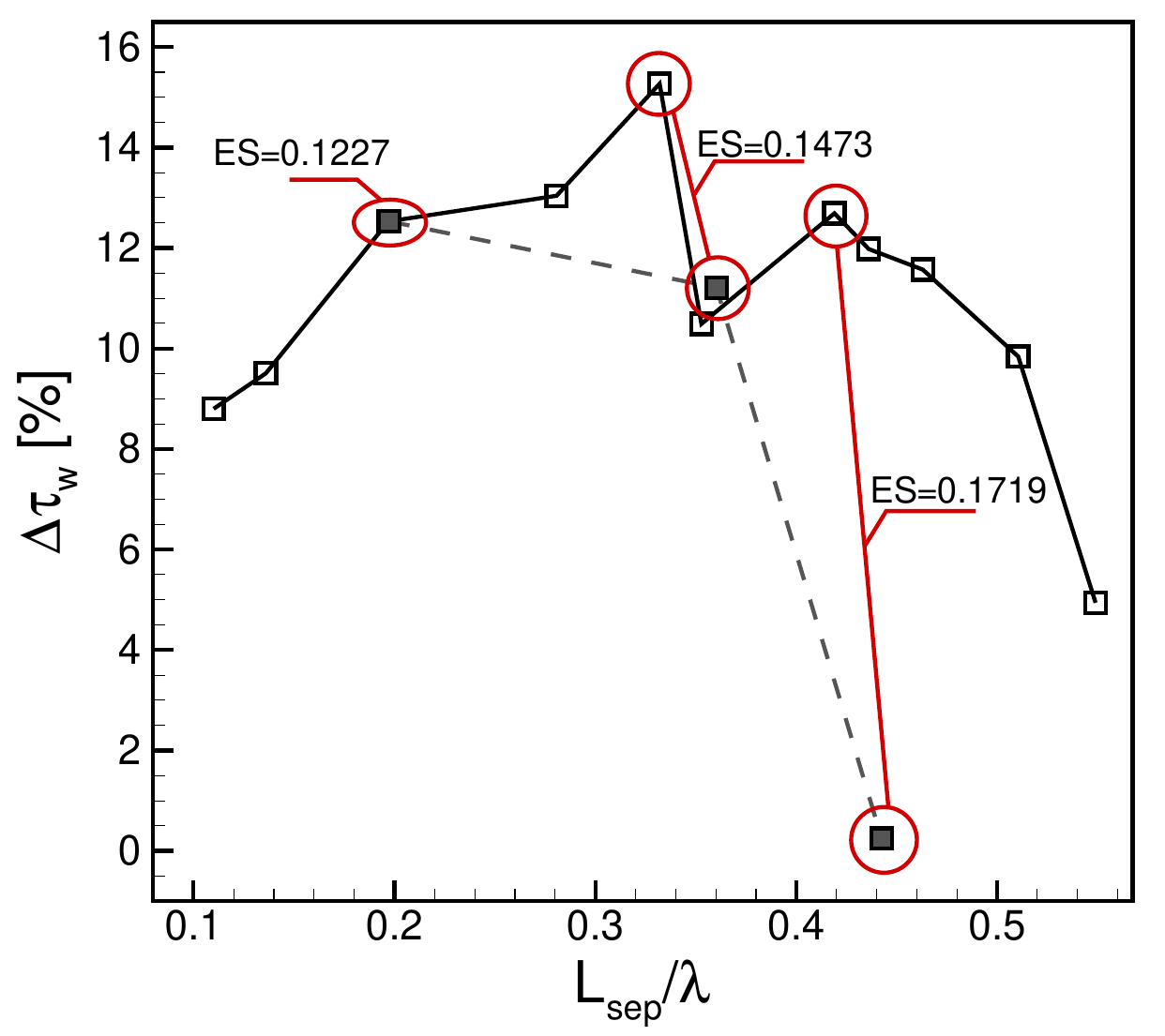}
    \caption{}
    \end{subfigure}
    \caption{Separation bubble length as a function of effective slope (a) wall-shear growth at $x=1.0$ m as a function of separation bubble length (b).}
    \label{fig:dtauw-es-sep}
\end{figure}

Figure \ref{fig:dtauw-es-sep} presents $L_{sep}/\lambda=f(ES)$ (a) and $\Delta\tau_w=f(L_{sep}/\lambda)$ (b) for cases with variable $A_0$ at constant $N_{\lambda}=5$ (solid line) and cases with variable $N_{\lambda}$ at constant $A_0=1.0$ (dashed line and grey symbols). It can be observed that for cases with identical $ES$ (outlined with red colour), the separation bubble length is comparable but the wall-shear stress enhancement at $x=1.0$ m (downstream of corrugation at the flat plate) differs significantly (see Fig.\ref{fig:effective-slope-2}), especially for $ES\geq0.1719$. On the other hand, cases with identical or comparable $L_{sep}/\lambda$ exhibit very similar levels of $\Delta\tau_w$ (see Fig. \ref{fig:dtauw-es-sep}b). This is especially true for $L_{sep}/\lambda\leq0.35$. The results characterised by $L_{sep}/\lambda = 0.2$ and $L_{sep}/\lambda \approx 0.35$ exhibit a very similar $\Delta\tau_w = 12.7$\% and $\Delta\tau_w\approx 11$\% respectively. On the other hand, there are two cases with $L_{sep}/\lambda \approx 0.44$ (and identical $ES=0.1719$) giving either marginal ($\Delta\tau_w=0.25$\%) or huge ($\Delta\tau_w=12$\%) growth in wall-shear stress. This might suggest that the separation bubble length $L_{sep}/\lambda$ is not a representative parameter that would potentially provide the information about $\Delta \tau_w$ change downstream the wavy wall. 

To sum up, it seems that $ES$ is not the only parameter determining an enhancement of the wall-shear stress due to the wavy wall. Bearing in mind that the amplitude modulation effect becomes enhanced with growing $Re$ \cite{drozdz2021effect}, one may expect that also more high-speed zones will emerge, and the separation point will move downstream. The results already discussed indicate that the wavy wall crests act in a similar manner as growing $Re$ in APG, namely, high-speed zones are generated above crests causing an enhanced production of small-scale spanwise vortices and the convection velocity of such structures decelerates slower than the mean flow on the downhill side of each crest. Since the convection of the spanwise small-scale vortices is higher, it generates strong sweeping motion in the troughs which, in result, enhances $\tau_w$ on the uphill side of the next crests. To obtain a greater production of the small-scale turbulence in high-speed zones and so, the higher wall-shear stress, one needs a large enough period of the corrugation. As suggested in \cite{drozdz2021effective}, the length of such a period should be of the order of the boundary layer thickness. From the so-far analysis of the separation bubble length change in troughs, one can conclude that the optimal geometry of the wavy wall (that ensures the highest growth in $\tau_w$) is that for which the flow is maintained in such a way that the detachment in the troughs does not exceed $L_{sep}/\lambda\approx0.3$. This observation is consistent with the work of Nugroho et al. \cite{nugroho_monty_utama_ganapathisubramani_hutchins_2021}, who stated that for condition where  $\lambda \approx \delta$, $ES$ is sufficiently low such that flow separations were minimised. This may suggest that the condition is desirable to obtain also the highest increase in the convection velocity of spanwise small-scale vortices with respect to the mean flow. On the other hand, this condition may be specific for a given $Re$ and APG. As shown in Table \ref{Tab:effective-slope-retau}, $ES$ that ensures the highest growth in $\tau_w$ decreases with the Reynolds number.

\begin{figure}[h!] 
    \centering
    \begin{subfigure}{0.99\linewidth}
    \includegraphics[trim={0 0 0 0}, clip, width=\linewidth]{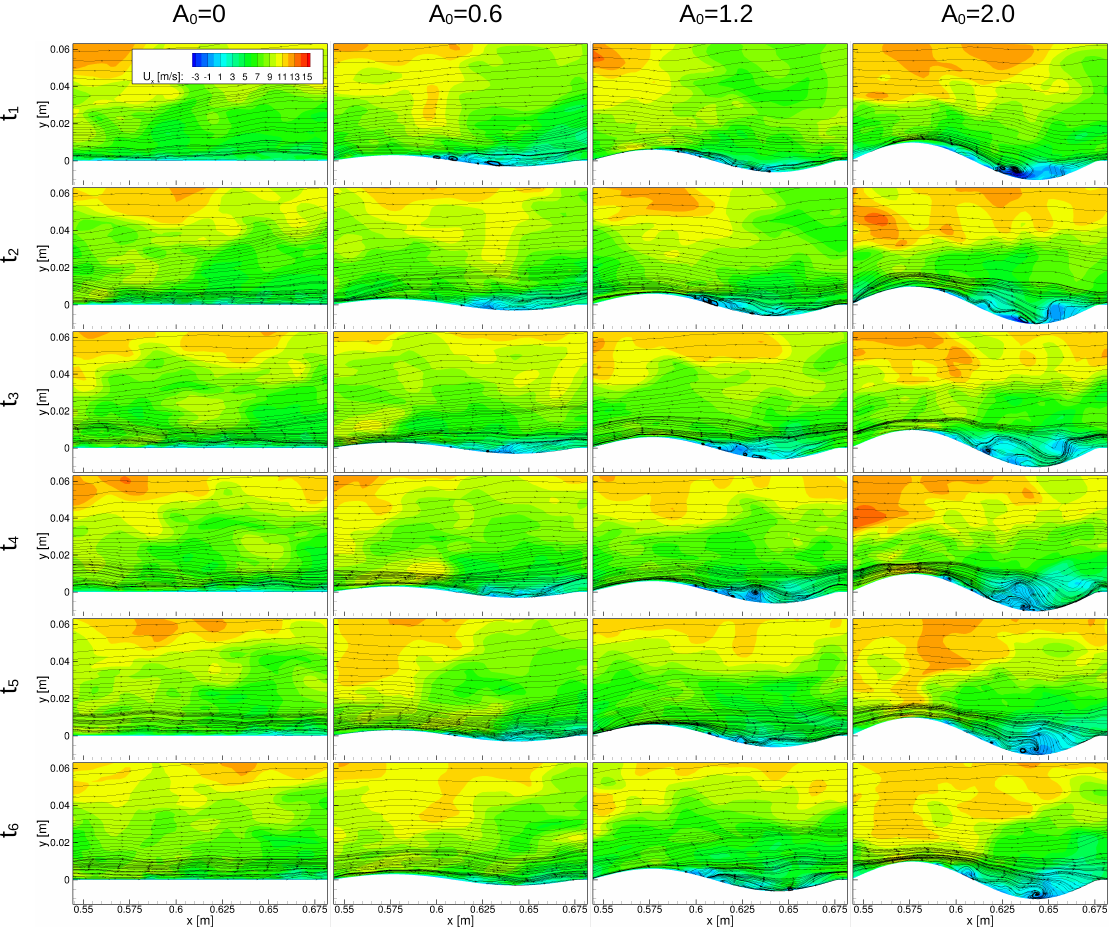}
    \end{subfigure}
    \caption{Instantaneous $U_x$ maps with streamlines for four $A_0$ at constant $N_{\lambda}=5$ and six time-instants $t_1-t_6$. Interval between each time-instant is equal to $\Delta t = 0.004$ s.}
    \label{fig:xvel-evolution}
\end{figure}

In Fig.~\ref{fig:xvel-evolution}, the solutions at four different instants are plotted to provide an even more detailed insight into the effect of the corrugations on the near-wall region. A zoom of the domain showing a single period (fifth/last period) is presented. Although there are some clear changes over time in the external flow, the analysis is focused on the near wall region. The dark blue colour and streamlines, superimposed to the contour plot, indicate regions with negative velocities that are in particular seen for larger amplitudes ($A_0>1.2$) within the waviness troughs and so, identifying in turn separation and reattachment zones. The recirculation region becomes larger with growing $A_0$ and for the largest amplitude ($A_0=2.0$) it occupies the streamwise distance wider than $\lambda/2$. This effect is not seen for the lower amplitudes. Worth noting is that the size of the separation bubble changes in time, which can be observed for instance for $A_0=1.2$, as it is relatively small at $t_1$ and becomes larger in time. Also, for $A_0=1.2$ at $t_2-t_4$ a strong sweeping motion, which prevents the downstream separation, can be seen just before the trough minimum. This is generated by the spanwise structures visible in Fig. \ref{fig:q-param-2}. Not less interesting is that if statistically-converged flow field of the case $A_0=0.6$ is analysed, no TBL detachment is observed (see Figs \ref{fig:effect-amplitudes-8} and \ref{fig:effect-amplitudes-4}). However, when looking at the instantaneous flow field (Fig. \ref{fig:xvel-evolution}), zones with negative velocities emerge ($A_0=0.6$). For large enough $A_0$, regions with a substantially enhanced velocity (for $y>0.03$ m) are seen, especially at $x=0.575$ m, which are generated by the strong local FPG caused by the presence of waviness. This effect is especially visible for $A_0=2.0$ so for the largest amplitude of corrugation. 
%

%%%%%%%%%%%%%%%%%%%%%%%%%%%%%%%%%%%%%%%%%%%%%%%%%%%%%%%%%%%%%%%%%%%%%%%%%%%%%%%%%%%%%%%%%%%%%%%%%%%
\section{Conclusions \label{conclusions}}
%%%%%%%%%%%%%%%%%%%%%%%%%%%%%%%%%%%%%%%%%%%%%%%%%%%%%%%%%%%%%%%%%%%%%%%%%%%%%%%%%%%%%%%%%%%%%%%%%%%
This study investigates (using LES) the effect of different geometries of transverse wall corrugation, in particular, the effect of such parameters as $A_0$ and $N_{\lambda}$ on the change in the wall-shear stress and the flow field under APG conditions. The key objective was to determine such a combination of $A_0$ and $N_{\lambda}$ leading to the highest possible enhancement of $\tau_w$ with respect to the configuration with the flat plate that should ensure the maximum postponement of TBL separation. LES simulations have been conducted for $U_{\infty}=15$ m/s which corresponds to relatively high $Re_{\tau}=2500$. 

The highest growth in the wall-shear with respect to the flow at the flat plate was found for $A_0=1.2$ and $N_{\lambda}=5$ which corresponds to $ES=0.1473$. For that case, the length of a waviness period is nearly of the order of the boundary layer thickness. An increase in $\Delta\tau_w$ is generally seen when $ES$ grows up to $0.15$ (so, within the so-called waviness regime of the surface corrugation). When $ES$ exceeds $0.15$ (so, when transitional waviness/roughness regime begins), $\Delta\tau_w$ starts decreasing. 

The present and previous studies showed that $ES$ ensuring the highest growth in $\Delta\tau_w$ seems to decrease asymptotically to a certain value with increasing Reynolds number while $A^+$ increases with Reynolds number. This may suggest that the corrugation amplitude should be of the order of the size of the near-wall region ($y^+\approx150$) that would be observed for the same flow case without corrugation, i.e. on the flat plate. 

A detailed insight into the flow field over the transverse surface waviness region has led to the following conclusions. The transverse surface waviness has a significant impact on the near-wall velocity gradient and consequently on the wall-shear stress, both, along the wavy wall and, what is more important, in the region further downstream. This is accompanied by the enhancement in TKE with the growth in either $A$ or $N_{\lambda}$. This change in TKE can be better described using $ES$ rather than $A$ or $N_{\lambda}$. Namely, the cases which are significantly diversified by $A$ and $N_{\lambda}$ but with identical $ES$ exhibit almost identical TKE profiles.
    
Local FPG and APG regions along the corrugation generate oscillations of $\tau_w$. Consequently, for large enough $A_0$ or $N_{\lambda}$, the separation inside the waviness trough is likely to occur. This effect is particularly evident for $A_0>0.8$ and $N_{\lambda}>5$, and it seems not to be correlated with $ES$. Moreover, as $A_0$ and $N_{\lambda}$ increase, the point of separation within each period of a waviness moves upstream and thus, the separation bubble length grows.

The presence of the waviness crests affects the flow similarly as large-scale motions at high-$Re$ TBL through the amplitude modulation mechanism \cite{mathis2009comparison}. Generation of high-speed zones by the crests causes the enhanced production of small-scale spanwise vortices with convection velocity decreasing slower than the mean flow on the downhill side of the crest. As a result, a stronger sweeping motion in the troughs is expected, which is responsible for the enhancement of $\tau_w$.
    
The presence of the waviness triggers the production of the longitudinal structures. It happens when the flow approaches each crest of the corrugation. Just downstream each crest, the detachment starts to occur and the longitudinal vortices break down and become transformed into spanwise structures, reminiscent of a hairpin. This effect is more pronounced for greater values of $A$ and it may be attributed to the postponement of the separation downstream of the waviness.

Although the results obtained in the present and previous works provide a general picture of the $\tau_w$ enhancement process through the wavy wall for a wide range of $Re_{\tau}$ (\tab{Tab:effective-slope-retau}), more results for even higher values of $Re_{\tau}$ as well as for different pressure gradient histories are of interest and so are recommended for future work.

\section{Acknowledgements}

The investigation was supported by the National Science Centre under Grant No. UMO-2020/39/B/ST8/01449. The simulations were carried out using the PL-Grid computer infrastructure.

\newpage
\bibliographystyle{ieeetr}
\bibliography{bibliography}

\begin{thebibliography}{100}

\bibitem{kaminski2024numerical}
P.~Kami{\'n}ski, P.~Niegodajew, A.~Dr{\'o}{\.z}d{\.z}, V.~Sokolenko, A.~Tyliszczak, and W.~Elsner, ``Numerical analysis of novel wavy wall based control of turbulent boundary layer separation,'' {\em Aerospace Science and Technology}, p.~109167, 2024.

\bibitem{lee2004investigation}
T.~Lee and P.~Gerontakos, ``Investigation of flow over an oscillating airfoil,'' {\em Journal of Fluid Mechanics}, vol.~512, pp.~313--341, 2004.

\bibitem{tanarro2020effect}
{\'A}.~Tanarro, R.~Vinuesa, and P.~Schlatter, ``Effect of adverse pressure gradients on turbulent wing boundary layers,'' {\em Journal of Fluid Mechanics}, vol.~883, p.~A8, 2020.

\bibitem{vila2020separating}
C.~S. Vila, R.~Vinuesa, S.~Discetti, A.~Ianiro, P.~Schlatter, and R.~{\"O}rl{\"u}, ``Separating adverse-pressure-gradient and {R}eynolds-number effects in turbulent boundary layers,'' {\em Physical Review Fluids}, vol.~5, no.~6, p.~064609, 2020.

\bibitem{azad1996turbulent}
R.~S. Azad, ``Turbulent flow in a conical diffuser: A review,'' {\em Experimental {T}hermal and {F}luid {S}cience}, vol.~13, no.~4, pp.~318--337, 1996.

\bibitem{apsley2000advanced}
D.~Apsley and M.~Leschziner, ``Advanced turbulence modelling of separated flow in a diffuser,'' {\em Flow, {T}urbulence and {C}ombustion}, vol.~63, pp.~81--112, 2000.

\bibitem{salehi2017computation}
S.~Salehi, M.~Raisee, and M.~Cervantes, ``Computation of developing turbulent flow through a straight asymmetric diffuser with moderate adverse pressure gradient,'' {\em Journal of Applied Fluid Mechanics}, vol.~10, no.~4, pp.~1029--1043, 2017.

\bibitem{yadegari2020numerical}
M.~Yadegari and A.~B. Khoshnevis, ``Numerical study of the effects of adverse pressure gradient parameter, turning angle and curvature ratio on turbulent flow in 3d turning curved rectangular diffusers using entropy generation analysis,'' {\em The European Physical Journal Plus}, vol.~135, no.~7, p.~548, 2020.

\bibitem{bons2005critical}
J.~Bons, ``A critical assessment of {R}eynolds analogy for turbine flows,'' {\em ASME Journal of Heat and Mass Transfer}, vol.~127, no.~5, pp.~472--485, 2005.

\bibitem{bons2003effect}
J.~P. Bons and S.~T. McClain, ``The effect of real turbine roughness with pressure gradient on heat transfer,'' in {\em Turbo Expo: Power for Land, Sea, and Air}, vol.~36886, pp.~611--622, 2003.

\bibitem{goyal2017experimental}
R.~Goyal, B.~K. Gandhi, and M.~J. Cervantes, ``Experimental study of mitigation of a spiral vortex breakdown at high {R}eynolds number under an adverse pressure gradient,'' {\em Physics of Fluids}, vol.~29, no.~10, 2017.

\bibitem{nagano1993effects}
Y.~Nagano, M.~Tagawa, and T.~Tsuji, ``Effects of adverse pressure gradients on mean flows and turbulence statistics in a boundary layer,'' in {\em Turbulent Shear Flows 8: Selected Papers from the Eighth International Symposium on Turbulent Shear Flows, Munich, Germany, September 9--11, 1991}, pp.~7--21, Springer, 1993.

\bibitem{kitsios2017direct}
V.~Kitsios, A.~Sekimoto, C.~Atkinson, J.~A. Sillero, G.~Borrell, A.~G. Gungor, J.~Jim{\'e}nez, and J.~Soria, ``Direct numerical simulation of a self-similar adverse pressure gradient turbulent boundary layer at the verge of separation,'' {\em Journal of Fluid Mechanics}, vol.~829, pp.~392--419, 2017.

\bibitem{peterson2019control}
C.~J. Peterson, B.~Vukasinovic, and A.~Glezer, ``Control of a closed separation domain in adverse pressure gradient over a curved surface,'' in {\em AIAA Scitech 2019 Forum}, p.~1900, 2019.

\bibitem{gad1991separation}
M.~Gad-el Hak and D.~M. Bushnell, ``Separation control,'' 1991.

\bibitem{joshi2016review}
S.~N. Joshi and Y.~S. Gujarathi, ``A review on active and passive flow control techniques,'' {\em International Journal on Recent Technologies in Mechanical and Electrical Engineering}, vol.~3, no.~4, pp.~1--6, 2016.

\bibitem{ashill2005review}
P.~Ashill, J.~Fulker, and K.~Hackett, ``A review of recent developments in flow control,'' {\em The Aeronautical Journal}, vol.~109, no.~1095, pp.~205--232, 2005.

\bibitem{you2008active}
D.~You and P.~Moin, ``Active control of flow separation over an airfoil using synthetic jets,'' {\em Journal of Fluids and Structures}, vol.~24, no.~8, pp.~1349--1357, 2008.

\bibitem{moghaddam2017active}
T.~Moghaddam and N.~B. Neishabouri, ``On the active and passive flow separation control techniques over airfoils,'' in {\em IOP Conference Series: Materials Science and Engineering}, vol.~248, p.~012009, IOP Publishing, 2017.

\bibitem{jahanmiri2010active}
M.~Jahanmiri, ``Active flow control: a review,'' 2010.

\bibitem{yoon2006drag}
H.~S. Yoon, O.~A. El-Samni, and H.~H. Chun, ``Drag reduction in turbulent channel flow with periodically arrayed heating and cooling strips,'' {\em Physics of Fluids}, vol.~18, no.~2, p.~025104, 2006.

\bibitem{harwigsson1996environmentally}
I.~Harwigsson and M.~Hellsten, ``Environmentally acceptable drag-reducing surfactants for district heating and cooling,'' {\em Journal of the American Oil Chemists' Society}, vol.~73, no.~7, pp.~921--928, 1996.

\bibitem{quadrio2011drag}
M.~Quadrio, ``Drag reduction in turbulent boundary layers by in-plane wall motion,'' {\em Philosophical Transactions of the Royal Society A: Mathematical, Physical and Engineering Sciences}, vol.~369, no.~1940, pp.~1428--1442, 2011.

\bibitem{leschziner2020friction}
M.~A. Leschziner, ``Friction-drag reduction by transverse wall motion--a review,'' {\em Journal of Mechanics}, vol.~36, no.~5, pp.~649--663, 2020.

\bibitem{vernet2018flow}
J.~A. Vernet, R.~{\"O}rl{\"u}, and P.~H. Alfredsson, ``Flow separation control behind a cylindrical bump using dielectric-barrier-discharge vortex generator plasma actuators,'' {\em Journal of Fluid Mechanics}, vol.~835, pp.~852--879, 2018.

\bibitem{vernet2018plasma}
J.~A. Vernet, R.~{\"O}rl{\"u}, D.~S{\"o}derblom, P.~Elofsson, and P.~H. Alfredsson, ``Plasma streamwise vortex generators for flow separation control on trucks: a proof-of-concept experiment,'' {\em Flow, {T}urbulence and {C}ombustion}, vol.~100, pp.~1101--1109, 2018.

\bibitem{vernet2015separation}
J.~A. Vernet, R.~{\"O}rl{\"u}, and P.~H. Alfredsson, ``Separation control by means of plasma actuation on a half cylinder approached by a turbulent boundary layer,'' {\em Journal of Wind Engineering and Industrial Aerodynamics}, vol.~145, pp.~318--326, 2015.

\bibitem{lorenz2006spinning}
R.~D. Lorenz, {\em Spinning flight: dynamics of frisbees, boomerangs, samaras, and skipping stones}.
\newblock Springer, 2006.

\bibitem{choi2006mechanism}
J.~Choi, W.-P. Jeon, and H.~Choi, ``Mechanism of drag reduction by dimples on a sphere,'' {\em Physics of Fluids}, vol.~18, no.~4, 2006.

\bibitem{tay2018drag}
J.~Tay and T.~T. Lim, ``Drag reduction with teardrop-shaped dimples,'' in {\em 2018 Flow Control Conference}, p.~3528, 2018.

\bibitem{tay2015mechanics}
C.~Tay, B.~Khoo, and Y.~Chew, ``Mechanics of drag reduction by shallow dimples in channel flow,'' {\em Physics of Fluids}, vol.~27, no.~3, 2015.

\bibitem{gattere2022dimples}
F.~Gattere, A.~Chiarini, and M.~Quadrio, ``Dimples for skin-friction drag reduction: status and perspectives,'' {\em Fluids}, vol.~7, no.~7, p.~240, 2022.

\bibitem{aoki2012mechanism}
K.~Aoki, K.~Muto, and H.~Okanaga, ``Mechanism of drag reduction by dimple structures on a sphere,'' {\em Journal of Fluid Science and Technology}, vol.~7, no.~1, pp.~1--10, 2012.

\bibitem{bearman1976golf}
P.~W. Bearman and J.~K. Harvey, ``Golf ball aerodynamics,'' {\em Aeronautical Quarterly}, vol.~27, no.~2, pp.~112--122, 1976.

\bibitem{veldhuis2009drag}
L.~Veldhuis and E.~Vervoort, ``Drag effect of a dented surface in a turbulent flow,'' in {\em 27th AIAA Applied Aerodynamics Conference}, p.~3950, 2009.

\bibitem{tay2011determining}
C.~Tay, ``Determining the effect of dimples on drag in a turbulent channel flow,'' in {\em 49th AIAA Aerospace Sciences Meeting Including the New Horizons Forum and Aerospace Exposition}, p.~682, 2011.

\bibitem{tay2019drag}
J.~Tay, T.~T. Lim, and B.~C. Khoo, ``Drag reduction with diamond-shaped dimples,'' in {\em AIAA Aviation 2019 Forum}, p.~3296, 2019.

\bibitem{koike2004research}
M.~Koike, T.~Nagayoshi, and N.~Hamamoto, ``Research on aerodynamic drag reduction by vortex generators,'' {\em Mitsubishi motors technical review}, vol.~16, pp.~11--16, 2004.

\bibitem{aider2010drag}
J.-L. Aider, J.-F. Beaudoin, and J.~E. Wesfreid, ``Drag and lift reduction of a 3d bluff-body using active vortex generators,'' {\em Experiments in Fluids}, vol.~48, pp.~771--789, 2010.

\bibitem{seshagiri2009effects}
A.~Seshagiri, E.~Cooper, and L.~W. Traub, ``Effects of vortex generators on an airfoil at low {R}eynolds numbers,'' {\em Journal of Aircraft}, vol.~46, no.~1, pp.~116--122, 2009.

\bibitem{belamadi2016aerodynamic}
R.~Belamadi, A.~Djemili, A.~Ilinca, and R.~Mdouki, ``Aerodynamic performance analysis of slotted airfoils for application to wind turbine blades,'' {\em Journal of Wind Engineering and Industrial Aerodynamics}, vol.~151, pp.~79--99, 2016.

\bibitem{coder2020design}
J.~G. Coder and D.~M. Somers, ``Design of a slotted, natural-laminar-flow airfoil for commercial transport applications,'' {\em Aerospace Science and Technology}, vol.~106, p.~106217, 2020.

\bibitem{whitman2006experimental}
N.~Whitman, R.~Sparks, S.~Ali, and J.~Ashworth, ``Experimental investigation of slotted airfoil performance with modified slot configurations,'' in {\em 24th AIAA Applied Aerodynamics Conference}, p.~3481, 2006.

\bibitem{casalino2019aeroacoustic}
D.~Casalino, F.~Avallone, I.~Gonzalez-Martino, and D.~Ragni, ``Aeroacoustic study of a wavy stator leading edge in a realistic fan/ogv stage,'' {\em Journal of Sound and Vibration}, vol.~442, pp.~138--154, 2019.

\bibitem{teruna2022numerical}
C.~Teruna, L.~Rego, D.~Casalino, D.~Ragni, and F.~Avallone, ``A numerical study on aircraft noise mitigation using porous stator concepts,'' {\em Aerospace}, vol.~9, no.~2, p.~70, 2022.

\bibitem{WANG2018101}
Y.~Wang, G.~Li, S.~Shen, D.~Huang, and Z.~Zheng, ``Influence of an off-surface small structure on the flow control effect on horizontal axis wind turbine at different relative inflow angles,'' {\em Energy}, vol.~160, pp.~101--121, 2018.

\bibitem{mostafa2022quantitative}
W.~Mostafa, A.~Abdelsamie, M.~Sedrak, D.~Th{\'e}venin, and M.~H. Mohamed, ``Quantitative impact of a micro-cylinder as a passive flow control on a horizontal axis wind turbine performance,'' {\em Energy}, vol.~244, p.~122654, 2022.

\bibitem{wang2023wake}
J.-S. Wang, J.~Wu, and J.-J. Wang, ``Wake-triggered secondary vortices over a cylinder/airfoil configuration,'' {\em Experiments in Fluids}, vol.~64, no.~1, p.~6, 2023.

\bibitem{smith2001performance}
M.~Smith, N.~Komerath, R.~Ames, O.~Wong, and J.~Pearson, ``Performance analysis of a wing with multiple winglets,'' in {\em 19th AIAA Applied Aerodynamics Conference}, p.~2407, 2001.

\bibitem{la2004induced}
U.~La~Roche and H.~L. La~Roche, ``Induced drag reduction using multiple winglets, looking beyond the prandtl-munk linear model,'' in {\em 2nd AIAA Flow Control Conference}, p.~2120, 2004.

\bibitem{guerrero2012biomimetic}
J.~E. Guerrero, D.~Maestro, and A.~Bottaro, ``Biomimetic spiroid winglets for lift and drag control,'' {\em Comptes Rendus Mecanique}, vol.~340, no.~1-2, pp.~67--80, 2012.

\bibitem{wu2018experimental}
L.~Wu, Z.~Jiao, Y.~Song, C.~Liu, H.~Wang, and Y.~Yan, ``Experimental investigations on drag-reduction characteristics of bionic surface with water-trapping microstructures of fish scales,'' {\em Scientific Reports}, vol.~8, no.~1, p.~12186, 2018.

\bibitem{guerrero2020variable}
J.~E. Guerrero, M.~Sanguineti, and K.~Wittkowski, ``Variable cant angle winglets for improvement of aircraft flight performance,'' {\em Meccanica}, vol.~55, pp.~1917--1947, 2020.

\bibitem{dou2012bionic}
Z.~Dou, J.~Wang, and D.~Chen, ``Bionic research on fish scales for drag reduction,'' {\em Journal of Bionic Engineering}, vol.~9, no.~4, pp.~457--464, 2012.

\bibitem{chen2013biomimetic}
H.~Chen, F.~Rao, X.~Shang, D.~Zhang, and I.~Hagiwara, ``Biomimetic drag reduction study on herringbone riblets of bird feather,'' {\em Journal of Bionic Engineering}, vol.~10, no.~3, pp.~341--349, 2013.

\bibitem{schlieter2016mechanical}
A.~Schlieter, R.~Pflumm, I.~Shakhverdova, R.~Naraparaju, U.~Schulz, C.~Leyens, M.~Sch{\"u}tze, and W.~Reimers, ``Mechanical properties of shark-skin like structured surfaces for high-temperature applications,'' {\em Advanced Engineering Materials}, vol.~18, no.~5, pp.~688--702, 2016.

\bibitem{lang2017separation}
A.~W. Lang, E.~M. Jones, and F.~Afroz, ``Separation control over a grooved surface inspired by dolphin skin,'' {\em Bioinspiration \& Biomimetics}, vol.~12, no.~2, p.~026005, 2017.

\bibitem{lissaman1983low}
P.~Lissaman, ``Low-{R}eynolds-number airfoils,'' {\em Annual {R}eview of {F}luid {M}echanics}, vol.~15, no.~1, pp.~223--239, 1983.

\bibitem{mueller2003aerodynamics}
T.~J. Mueller and J.~D. DeLaurier, ``Aerodynamics of small vehicles,'' {\em Annual {R}eview of {F}luid {M}echanics}, vol.~35, no.~1, pp.~89--111, 2003.

\bibitem{mcmasters1979low}
J.~H. McMasters and M.~L. Henderson, ``Low-speed single-element airfoil synthesis,'' {\em NASA. Langley Res. Center The Sci. and Technol. of Low Speed and Motorless Flight, Pt. 1}, 1979.

\bibitem{aubertine2004parameters}
C.~D. Aubertine, J.~K. Eaton, and S.~Song, ``Parameters controlling roughness effects in a separating boundary layer,'' {\em International {J}ournal of {H}eat and {F}luid {F}low}, vol.~25, no.~3, pp.~444--450, 2004.

\bibitem{mejia2013wall}
R.~Mejia-Alvarez and K.~T. Christensen, ``Wall-parallel stereo particle-image velocimetry measurements in the roughness sublayer of turbulent flow overlying highly irregular roughness,'' {\em Physics of Fluids}, vol.~25, no.~11, 2013.

\bibitem{cherukat1998direct}
P.~Cherukat, Y.~Na, T.~Hanratty, and J.~McLaughlin, ``Direct numerical simulation of a fully developed turbulent flow over a wavy wall,'' {\em Theoretical and Computational Fluid Dynamics}, vol.~11, no.~2, pp.~109--134, 1998.

\bibitem{sun2018direct}
Z.~Sun, Y.~Zhu, Y.~Hu, and S.~Zhang, ``Direct numerical simulation of a fully developed compressible wall turbulence over a wavy wall,'' {\em Journal of Turbulence}, vol.~19, no.~1, pp.~72--105, 2018.

\bibitem{kuhn2010large}
S.~Kuhn, S.~Kenjere{\v{s}}, and P.~R. von Rohr, ``Large eddy simulations of wall heat transfer and coherent structures in mixed convection over a wavy wall,'' {\em International {J}ournal of {T}hermal {S}ciences}, vol.~49, no.~7, pp.~1209--1226, 2010.

\bibitem{tyson2013numerical}
C.~Tyson and N.~Sandham, ``Numerical simulation of fully-developed compressible flows over wavy surfaces,'' {\em International {J}ournal of {H}eat and {F}luid {F}low}, vol.~41, pp.~2--15, 2013.

\bibitem{kruse2006structure}
N.~Kruse and P.~R. Von~Rohr, ``Structure of turbulent heat flux in a flow over a heated wavy wall,'' {\em International {J}ournal of {H}eat and {M}ass {T}ransfer}, vol.~49, no.~19-20, pp.~3514--3529, 2006.

\bibitem{hamed2015turbulent}
A.~M. Hamed, A.~Kamdar, L.~Castillo, and L.~P. Chamorro, ``Turbulent boundary layer over 2d and 3d large-scale wavy walls,'' {\em Physics of Fluids}, vol.~27, no.~10, p.~106601, 2015.

\bibitem{elsner2022experimental}
W.~Elsner, A.~Dr{\'o}{\.z}d{\.z}, E.~Szymanek, A.~Tyliszczak, and P.~Niegodajew, ``Experimental and numerical studies of turbulent flows over two-dimensional and three-dimensional rough surfaces under an adverse pressure gradient,'' {\em Applied Mathematical Modelling}, vol.~106, pp.~549--566, 2022.

\bibitem{akselsen2020langmuir}
A.~H. Akselsen and S.~{\AA}. Ellingsen, ``Langmuir-type vortices in wall-bounded flows driven by a criss-cross wavy wall topography,'' {\em Journal of Fluid Mechanics}, vol.~900, p.~A19, 2020.

\bibitem{fernex2020actuation}
D.~Fernex, R.~Semaan, M.~Albers, P.~S. Meysonnat, W.~Schr{\"o}der, and B.~R. Noack, ``Actuation response model from sparse data for wall turbulence drag reduction,'' {\em Physical Review Fluids}, vol.~5, no.~7, p.~073901, 2020.

\bibitem{de1997direct}
V.~De~Angelis, P.~Lombardi, and S.~Banerjee, ``Direct numerical simulation of turbulent flow over a wavy wall,'' {\em Physics of Fluids}, vol.~9, no.~8, pp.~2429--2442, 1997.

\bibitem{koyama2007turbulence}
S.~Koyama, K.~Takashima, and Y.~Hagiwara, ``Turbulence modification in flow around a periodically deforming film,'' {\em Journal of Turbulence}, no.~8, p.~N19, 2007.

\bibitem{yoon2009effect}
H.~Yoon, O.~El-Samni, A.~Huynh, H.~Chun, H.~Kim, A.~Pham, and I.~Park, ``Effect of wave amplitude on turbulent flow in a wavy channel by direct numerical simulation,'' {\em Ocean Engineering}, vol.~36, no.~9-10, pp.~697--707, 2009.

\bibitem{fujii2011turbulence}
H.~Fujii, K.~Sakurai, T.~Nakano, and Y.~Hagiwara, ``Turbulence structure, friction drag and pressure drag due to turbulent flow over angled wavy surfaces,'' in {\em Seventh International Symposium on Turbulence and Shear Flow Phenomena}, Begel House Inc., 2011.

\bibitem{ghebali2017turbulent}
S.~Ghebali, S.~I. Chernyshenko, and M.~A. Leschziner, ``Can large-scale oblique undulations on a solid wall reduce the turbulent drag?,'' {\em Physics of Fluids}, vol.~29, no.~10, 2017.

\bibitem{segunda2018experimental}
V.~M. Segunda, S.~J. Ormiston, and M.~F. Tachie, ``Experimental and numerical investigation of developing turbulent flow over a wavy wall in a horizontal channel,'' {\em European Journal of Mechanics-B/Fluids}, vol.~68, pp.~128--143, 2018.

\bibitem{hamed2017turbulent}
A.~M. Hamed, L.~Castillo, and L.~P. Chamorro, ``Turbulent boundary layer response to large-scale wavy topographies,'' {\em Physics of Fluids}, vol.~29, no.~6, p.~065113, 2017.

\bibitem{drozdz2021effective}
A.~Dr{\'o}{\.z}d{\.z}, P.~Niegodajew, M.~Roma{\'n}czyk, V.~Sokolenko, and W.~Elsner, ``Effective use of the streamwise waviness in the control of turbulent separation,'' {\em Experimental {T}hermal and {F}luid {S}cience}, vol.~121, p.~110291, 2021.

\bibitem{mathis2009comparison}
R.~Mathis, J.~P. Monty, N.~Hutchins, and I.~Marusic, ``Comparison of large-scale amplitude modulation in turbulent boundary layers, pipes, and channel flows,'' {\em Physics of Fluids}, vol.~21, no.~11, p.~111703, 2009.

\bibitem{dogan2019quantification}
E.~Dogan, R.~{\"O}rl{\"u}, D.~Gatti, R.~Vinuesa, and P.~Schlatter, ``Quantification of amplitude modulation in wall-bounded turbulence,'' {\em Fluid Dynamics Research}, vol.~51, no.~1, p.~011408, 2019.

\bibitem{andreolli2023separating}
A.~Andreolli, D.~Gatti, R.~Vinuesa, R.~{\"O}rl{\"u}, and P.~Schlatter, ``Separating large-scale superposition and modulation in turbulent channels,'' {\em Journal of Fluid Mechanics}, vol.~958, p.~A37, 2023.

\bibitem{napoli2008effect}
E.~Napoli, V.~Armenio, and M.~De~Marchis, ``The effect of the slope of irregularly distributed roughness elements on turbulent wall-bounded flows,'' {\em Journal of Fluid Mechanics}, vol.~613, pp.~385--394, 2008.

\bibitem{schultz2009turbulent}
M.~P. Schultz and K.~A. Flack, ``Turbulent boundary layers on a systematically varied rough wall,'' {\em Physics of Fluids}, vol.~21, no.~1, 2009.

\bibitem{nugroho_monty_utama_ganapathisubramani_hutchins_2021}
B.~Nugroho, J.~P. Monty, I.~K. A.~P. Utama, B.~Ganapathisubramani, and N.~Hutchins, ``Non-$k$-type behaviour of roughness when in-plane wavelength approaches the boundary layer thickness,'' {\em Journal of Fluid Mechanics}, vol.~911, p.~A1, 2021.

\bibitem{de2016large}
M.~De~Marchis, ``Large eddy simulations of roughened channel flows: Estimation of the energy losses using the slope of the roughness,'' {\em Computers \& Fluids}, vol.~140, pp.~148--157, 2016.

\bibitem{forooghi2017toward}
P.~Forooghi, A.~Stroh, F.~Magagnato, S.~Jakirli{\'c}, and B.~Frohnapfel, ``Toward a universal roughness correlation,'' {\em Journal of Fluids Engineering}, vol.~139, no.~12, 2017.

\bibitem{Georgiadis2010}
N.~J. Georgiadis, D.~P. Rizzetta, and C.~Fureby, ``Large-eddy simulation: current capabilities, recommended practices, and future research,'' {\em AIAA Journal}, vol.~48, no.~8, pp.~1772--1784, 2010.

\bibitem{Zhiyin2015}
Y.~Zhiyin, ``Large-eddy simulation: Past, present and the future,'' {\em Chinese Journal of Aeronautics}, vol.~28, no.~1, pp.~11--24, 2015.

\bibitem{Geurts2019}
B.~J. Geurts, A.~Rouhi, and U.~Piomelli, ``Recent progress on reliability assessment of large-eddy simulation,'' {\em Journal of Fluids and Structures}, vol.~91, p.~102615, 2019.

\bibitem{nicoud1999subgrid}
F.~Nicoud and F.~Ducros, ``Subgrid-scale stress modelling based on the square of the velocity gradient tensor,'' {\em Flow, {T}urbulence and {C}ombustion}, vol.~62, no.~3, pp.~183--200, 1999.

\bibitem{fluent-theory-guide}
{\em Ansys Fluent Theory Guide}.
\newblock ANSYS Inc., 2021.

\bibitem{clauser1956turbulent}
F.~H. Clauser, ``The {T}urbulent {B}oundary {L}ayer,'' {\em Advances in {A}pplied {M}echanics}, vol.~4, pp.~1--51, 1956.

\bibitem{vinuesa2021high}
R.~Vinuesa, ``High-fidelity simulations in complex geometries: Towards better flow understanding and development of turbulence models,'' {\em Results in Engineering}, vol.~11, p.~100254, 2021.

\bibitem{pope2001turbulent}
S.~B. Pope, ``Turbulent flows,'' {\em Measurement Science and Technology}, vol.~12, no.~11, pp.~2020--2021, 2001.

\bibitem{hunt1988eddies}
J.~C. Hunt, A.~A. Wray, and P.~Moin, ``Eddies, streams, and convergence zones in turbulent flows,'' {\em Studying turbulence using numerical simulation databases, 2. Proceedings of the 1988 summer program}, 1988.

\bibitem{zhang2022numerical}
E.~Zhang, X.~Wang, and Q.~Liu, ``Numerical investigation on the temporal and spatial statistical characteristics of turbulent mass transfer above a two-dimensional wavy wall,'' {\em International {J}ournal of {H}eat and {M}ass {T}ransfer}, vol.~184, p.~122260, 2022.

\bibitem{zhang2021large}
E.~Zhang, X.~Wang, and Q.~Liu, ``Large-eddy simulation of turbulent flow over wavy wall with different wave steepness,'' in {\em E3S Web of Conferences}, vol.~299, p.~03012, EDP Sciences, 2021.

\bibitem{bobke_vinuesa_örlü_schlatter_2017}
A.~Bobke, R.~Vinuesa, R.~Örlü, and P.~Schlatter, ``History effects and near equilibrium in adverse-pressure-gradient turbulent boundary layers,'' {\em Journal of Fluid Mechanics}, vol.~820, p.~667–692, 2017.

\bibitem{NIEGODAJEW2019108456}
P.~Niegodajew, A.~Dróżdż, and W.~Elsner, ``A new approach for estimation of the skin friction in turbulent boundary layer under the adverse pressure gradient conditions,'' {\em International {J}ournal of {H}eat and {F}luid {F}low}, vol.~79, p.~108456, 2019.

\bibitem{drozdz2018passive}
A.~Dr{\'o}{\.z}d{\.z}, W.~Elsner, and D.~Sikorski, ``Passive skin friction control near turbulent separation--preliminary results,'' in {\em Journal of Physics: Conference Series}, vol.~1101, p.~012004, IOP Publishing, 2018.

\bibitem{buschmann2006recent}
M.~H. Buschmann and M.~Gad-el Hak, ``Recent developments in scaling of wall-bounded flows,'' {\em Progress in Aerospace Sciences}, vol.~42, no.~5-6, pp.~419--467, 2006.

\bibitem{zhang2021effects}
E.~Zhang, X.~Wang, and Q.~Liu, ``Effects of the spanwise heterogeneity of a three-dimensional wavy wall on momentum and scalar transport,'' {\em Physics of Fluids}, vol.~33, no.~5, p.~055116, 2021.

\bibitem{drozdz2023convection}
A.~Dr{\'o}{\.z}d{\.z}, P.~Niegodajew, M.~Roma{\'n}czyk, and W.~Elsner, ``Convection velocity in turbulent boundary layers under adverse pressure gradient,'' {\em Experimental {T}hermal and {F}luid {S}cience}, p.~110900, 2023.

\bibitem{SCHLATTER201475}
P.~Schlatter, Q.~Li, R.~Örlü, F.~Hussain, and D.~Henningson, ``On the near-wall vortical structures at moderate {R}eynolds numbers,'' {\em European Journal of Mechanics - B/Fluids}, vol.~48, pp.~75--93, 2014.

\bibitem{schoppa_hussain_2002}
W.~SCHOPPA and F.~HUSSAIN, ``Coherent structure generation in near-wall turbulence,'' {\em Journal of Fluid Mechanics}, vol.~453, p.~57–108, 2002.

\bibitem{drozdz2021effect}
A.~Dr{\'o}{\.z}d{\.z}, P.~Niegodajew, M.~Roma{\'n}czyk, and W.~Elsner, ``Effect of {R}eynolds number on turbulent boundary layer approaching separation,'' {\em Experimental {T}hermal and {F}luid {S}cience}, vol.~125, p.~110377, 2021.

\end{thebibliography}
%%%%%%%%%%%%%%%%%%%%%%%%%%%%%%%%%%%%%%%%%%%%%%%%%%%%%%%%%%%%%%%%%%%%%%%%%%%%%%%%%%%%%%%%%%%%%%%%%%%
%

\clearpage
\thispagestyle{empty}
\setcounter{page}{0}

\begin{singlespace}

\clearpage
%%%%%%%%\clearpage%%%%%%%%%%%%%%%%%%%%%%%%%%%%%%%%%%%%%

\end{singlespace}

\end{document}